\documentclass[pre, superscriptaddress]{revtex4}
\usepackage{amsmath}                   
\usepackage{amsfonts}                  
\usepackage{epsfig}                    
\usepackage{latexsym}

\begin{document}
\title{Mechanical unfolding of a homopolymer globule studied by self-consistent field modelling}

\author{Alexey A. Polotsky}
\email{alexey.polotsky@gmail.com}
\affiliation{Institute of Macromolecular Compounds, Russian Academy of Sciences 31 Bolshoy pr., 199004 St.-Petersburg, Russia}

\author{Marat I. Charlaganov}
\affiliation{Laboratory of Physical Chemistry and Colloid Science, Wageningen University, The Netherlands}

\author{Frans A. M. Leermakers}
\affiliation{Laboratory of Physical Chemistry and Colloid Science, Wageningen University, The Netherlands}

\author{Mohamed Daoud}
\affiliation{Service de Physique de l'Etat Condens\'e CEA Saclay, 91191 Gif-sur-Yvette Cedex, France}

\author{Oleg V. Borisov}
\affiliation{Institute of Macromolecular Compounds, Russian Academy of Sciences 31 Bolshoy pr., 199004 St.-Petersburg, Russia}
\affiliation{Institut Pluridisciplinaire de Recherche sur l'Environnement et les Mat\'eriaux, UMR 5254 CNRS/UPPA, Pau, France}

\author{Tatiana M. Birshtein}
\affiliation{Institute of Macromolecular Compounds, Russian Academy of Sciences 31 Bolshoy pr., 199004 St.-Petersburg, Russia}

\begin{abstract}
We present results of numerical Self-Consistent Field (SCF) calculations for the equilibrium 
mechanical unfolding of a globule formed by a single flexible polymer chain collapsed in a poor solvent. 
In accordance with earlier scaling theory and stochastic dynamics simulations findings
we have identified three regimes of extensional deformation: 
(i)~a linear response regime characterized by a weakly elongated (ellipsoidal) shape of the globule at small deformations; 
(ii)~a tadpole structure with a globular ``head'' co-existing with a stretched ``tail'' at intermediate ranges of deformations and 
(iii)~an uniformly stretched chain at strong extensions. 
The conformational transition from the tadpole to the stretched chain is accompanied by an abrupt unfolding of the depleted globular head and a corresponding jump-wide drop in the intra-chain tension. The unfolding-refolding cycle demonstrates a hysteresis loop in the vicinity of the transition point. 
These three regimes of deformation, as well as the first-order like transition between the tadpole and the stretched chain conformations, 
can be experimentally observable provided that the number of monomer units in the chain is large and/or the solvent quality is sufficiently poor. For short chains, on the other hand, at moderately poor solvent strength conditions the unfolding transition is continuous. Upon an increase in the imposed end-to-end distance the extended globule retains a longitudinally uniform shape at any degree of deformation. In all cases the system exhibits a negative extensional modulus in the intermediate range of deformations. We anticipate that predictions of patterns in force-deformation curves for polymer molecules in poor solvent can be observed in single molecule atomic spectroscopy experiments.  
\end{abstract}

\maketitle


\section{Introduction} \label{Sec:Intro}
The particular interest in the globular state of individual macromolecules (collapsed in poor solvent) 
and in the conformational collapse-to-swelling or unfolding transitions in individual polymer chains is 
motivated by the existing physical analogy between globules of synthetic polymers stabilized by solvophobic 
(attractive) interactions between the monomer units in poor solvents and the compact structures found for biopolymers, e.g., 
for globular proteins \cite{PtitsynFinkelshtein:2002}.

Despite this profound analogy and the fact that these objects share the same name, homopolymer globules differ markedly  
from protein ones. A globular protein, which is a copolymer with a large number of different amino acid residues, 
typically has a unique intra-molecular structure being a  ``aperiodic crystal''  \cite{Schroedinger:1944}. 
A globule of a flexible homopolymer can, in contrast, be better compared to a liquid droplet. 
In this case, the coil-globule transition is similar to the usual gas-liquid transition upon a decrease 
in the temperature (alternatively to the increase in the attraction between gas particles). 
The connectivity of the interacting monomers inside a chain (``linear memory'') changes the characteristics 
of this transition to some extent; the transition keeps its phase nature but now becomes a continuous second order phase transition. 
The conformational and thermodynamic characteristics of a globule stabilized by monomer-monomer attractions are functions of the strength 
of this attraction, or in other words, are determined by the solvent quality. The density (segment concentration) 
of the globule grows monotonously upon a worsening of the solvent quality from a very low value $\sim N^{-1/2}$ 
in the Gaussian coil to a value of order unity in a densely packed globule. 
The remainder of the volume of a globule is occupied by solvent molecules. 
In the protein globule, on the other hand, the whole inner space is filled by the polypeptide chain and the globule is nearly dry.

Recent developments in AFM force spectroscopy and optical tweezers techniques have made 
it possible to manipulate individual molecules and to subject them to mechanical force, for example, 
to stretch the chain by  extending the distance between 
its two ends~\cite{Rief:1997, Hugel:2001, Haupt:2002, Smith:1996, Kellermayer:1997} .

In essence, there are two possible scenarios for such extension. 
In the first case the governing parameter is the distance between two points in a macromolecule (for example between the two end segments) which is fixed to a specified value or changed with a given velocity. The observable in this case is the reaction, or restoring, force. In the second case the applied force is fixed and plays the role of the control parameter. The observable in this case is the average end-to-end distance. In both scenarios a force-extension relation (or force-extension curve) is obtained which for finite chains not necessarily are identical. Note that in experiments with individual macromolecules the relevant distances are in the nanometer (nm) domain, whereas the force is in the piconewton (pN) range. In a recent review of Skvortsov \emph{et al.}~\cite{Skvortsov:2009} it was demonstrated that the constant extension ensemble (first scenario) which is mostly used in experiments, leads to a remarkable reach deformation behaviour.

There exists a large number of experimental works devoted to the extension of globular proteins. The objectives of such studies are to find, e.g., ``weak spots'' in a globule structure, or to discover a possible folding pathway. It has been shown in experiments on unfolding of proteins~\cite{Forman:2005} that the force versus deformation curves may exhibit quite complex patterns and are essentially non-monotonic.

\begin{figure}[t] 
  \begin{center}
  \includegraphics[width=10cm]{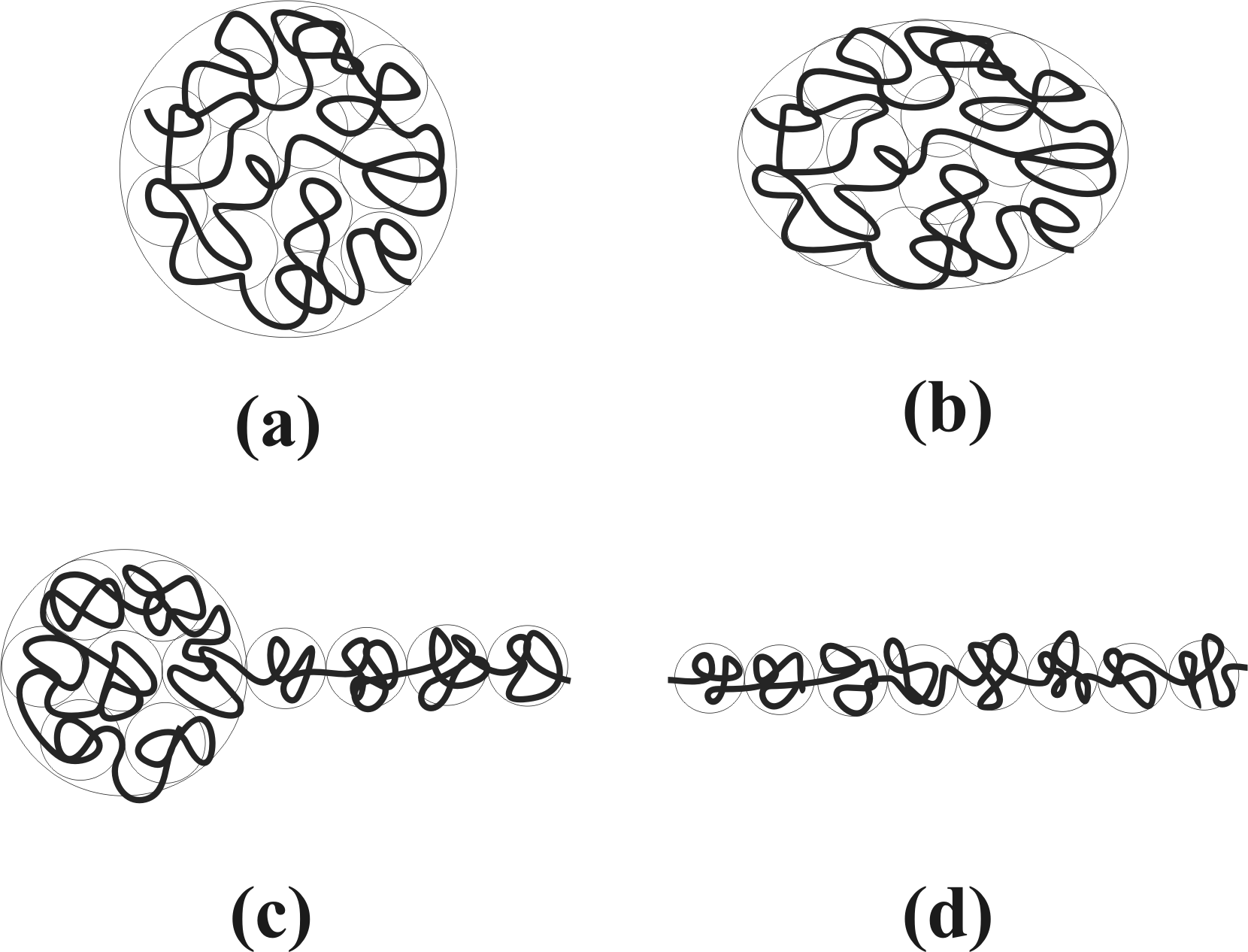}
  \end{center}
  \caption{Unperturbed spherical  (a) and ellipsoidal (b) globules, tadpole (c), and uniformly stretched (d) conformations.}
  \label{fig:globules}
\end{figure}

The pioneering theory of unfolding homopolymer globules subjected to an extensional deformation, 
was proposed by Halperin and Zhulina~\cite{Halperin:1991:EL}. 
This theory envisions that a weak extensional deformation of a spherical globule, Figure~\ref{fig:globules}~(a), 
produces a prolate ellipsoid of increasing asymmetry, Figure~\ref{fig:globules}~(b). 
The reaction (restoring) force was predicted to grow linearly at this stage. 
Under a moderate extensional deformation, a coexistence between a collapsed globular core and an extended 
``tail'' takes place within a single macromolecule, Figure~\ref{fig:globules}~(c). 
The deformation in this ``tadpole'' regime is accompanied by a progressive unfolding of the globular 
core which occurs at an almost constant reaction force. This stage ends when a stretched string of thermal blobs is obtained, 
i.e. when the size of the tadpole's globular head becomes of the order of the thermal blob size, Figure~\ref{fig:globules}~(d).  
From this point onward the reaction force grows again with the following extension of the unfolded chain.

Later on, 
Cooke and Williams~\cite{Cooke:2003} demonstrated the existence of a first-order 
conformational transition in the stretching of a collapsed (dry) polymer: 
at certain extension the chain suddenly unravels from the tadpole conformation, 
Figure~\ref{fig:globules}~(c), to the open chain conformation, Figure~\ref{fig:globules}~(d). 
In the force-extension curve this transition appears as a discontinuous drop in the force. 
The collapsed "head" in the transition point contains $\sim N^{3/4}$ monomer units.
The discontinuous drop in the force disappears, however, in the thermodynamic limit of infinitely long chain, $N\rightarrow \infty$.
Similar unraveling transition was found by Craig and Terentiev~\cite{Craig:2005:1} 
who studied unfolding of globules made by semi-flexible polymers.

The analogy between a homopolymer globule and a liquid droplet has been mentioned already. 
This analogy is also important for understanding the globule unfolding. 
The chain stretching, that is, the progressive increase in the end-to-end distance of the chain, 
is similar to increasing the available volume in the liquid-gas transition. 
The role of the gas phase is played in this case by the extended tail drawn out of the globule. 
Gas-liquid phase coexistence in a certain range of volumes (in our case - in the certain range of the given end-to-end distances) 
or at a constant pressure (in our case - at a constant reaction force) 
is well-known~\cite{Landau:5}. 
In the case of the polymer globule two phases coexist in a single macromolecule. 
This behavior is closely related to the Rayleigh instability in a liquid droplet~\cite{Rayleigh:1882}. 
In the case of a polymer in poor solvent an additional connectivity constraint of the solvophobic monomers in the chain~\cite{Dobrynin:1996} 
comes into play. 
Similar force-deformation patterns have been predicted 
for the unfolding transition in globular structures of amphiphilic associating copolymers ~\cite{Borisov:1999}. 
The topological complexity, that is, the grafting of polymer chains onto a surface (a brush), introduces even more interesting 
features in the scenario of the unfolding transition~\cite{Halperin:1991:MM},~\cite{Klushin:1998}. 

The aim of the present study is to develop a general theory of the polymer globule deformation 
as a function of the degree of polymerization $N$ (number of monomer units in the chain) and the solvent quality. 
These two parameters determine the properties of the free unperturbed globule. 
Our work includes several stages. In the first part presented in this paper, 
we use the Scheutjens-Fleer self-consistent field (SF-SCF) lattice approach 
in its two-gradient version to study the evolution of the conformational and thermodynamic properties of a polymer globule 
upon a uniaxial extension (i.e. with an increase in the distance between the chain ends). 
We consider a wide range of $N$ and polymer-solvent interaction parameter values. In the second part, an analytical Flory-type theory for the unfolding of a globule is advanced. The analytical theory uses, as input parameters, the properties of an unperturbed globule found with the aid of SF-SCF approach. This theory allows us to go beyond the limits of the SF-SCF modeling (first of all, to consider the limit of very long chains) and to calculate phase diagrams of the deformed globule as well as the properties of the globule in the transition point. These issues related to the phase diagrams, however, will be presented elsewhere.

The remainder of the paper is organized as follows. In the section ``Model and method'' we introduce the SF-SCF approach. The results of the calculations are summarized in the section ``Results''. In the section ``Blob picture of globule deformation'' we discuss the obtained results in terms of the classical scaling theory by Halperin and Zhulina for the extension of a flexible linear polymer chain collapsed in a poor solvent, and this is followed by the Conclusions.

\section{Model and method} \label{sec:SCF}
\subsection{Theoretical background of the self-consistent field (SCF) approach} \label{sec:SCF:theor}
At the basis of the SCF approach is a mean-field free energy which is expressed as a functional of the volume fraction profiles 
(normalized concentrations) and the self-consistent field potentials. 
The minimization of this free energy leads for polymer chains to the Edwards diffusion differential equation~\cite{deGennes:1979}, which for an arbitrary coordinate system may be expressed as
\begin{equation} \label{eq:EDE}
  \frac{\partial G(\mathbf{r}', \mathbf{r};  n)}{\partial n} = \frac{a^2}{6} \nabla^2 G(\mathbf{r}', \mathbf{r}; n) -   \frac{u(\mathbf{r})}{k_BT} \, G(\mathbf{r}', \mathbf{r}; n)
\end{equation}
($k_B$ is the Boltzmann constant, $T$ is the absolute temperature). 

The Green's function  $G((\mathbf{r}',\mathbf{r}; n)$ used in Eq.~(\ref{eq:EDE}) is the statistical weight of a probe chain with the length $n$ having its ends fixed in the points $\mathbf{r}'$ and $\mathbf{r}$. The self-consistent potential $u(\mathbf{r})$ represents the surrounding of the chain and serves as an external field used in the Boltzmann equation to find the statistical weight for each chain conformation. Consequently, the Green's functions $G(\mathbf{r}', \mathbf{r}; n)$ that obey Eq.~(\ref{eq:EDE}) is related to the volume fraction profile of the polymer by a composition law: 
\begin{equation} \label{eq:phi_scf}
  \varphi(\mathbf{r}) = \frac{\sum_{\mathbf{r}'} \sum_{\mathbf{r}''} \sum_{n} G(\mathbf{r}', \mathbf{r};  n) 
  G(\mathbf{r}, \mathbf{r}'';  N-n)}{\sum_{\mathbf{r}'} \sum_{\mathbf{r}''} G(\mathbf{r}'', \mathbf{r}';  N)} .
\end{equation}
In our case the ends of the chain are ``pinned'' at the points $\mathbf{r}'$ and $\mathbf{r}''$, so, the usual summation over the end points position  should be omitted:
\begin{equation} \label{eq:phi_scf_pinned}
  \varphi(\mathbf{r}) \equiv \varphi(\mathbf{r};\mathbf{r}',\mathbf{r}'') = \frac{\sum_{n} G(\mathbf{r}', \mathbf{r};  n) G(\mathbf{r}, \mathbf{r}'';  N-n)}{G(\mathbf{r}'',
  \mathbf{r}';  N)} .
\end{equation}
The boundary conditions and incompressibility condition: $\varphi(\mathbf{r})+\varphi_S(\mathbf{r})=1$, 
where $\varphi_S(\mathbf{r})$ is the volume fraction of the monomeric solvent, provide constraints on the spatial solutions.  
The potential $u(\mathbf{r})$ is local (i.e. there are no long-range forces) and depends on the local volume fraction $\varphi(\mathbf{r})$
\begin{equation} \label{eq:u_scf_general}
  u(\mathbf{r}) = u[\varphi(\mathbf{r})] ,
\end{equation}
The explicit concentration dependence will be specified below (see Eq.~(\ref{eq:u_potetntial})). Equations (\ref{eq:EDE}), (\ref{eq:phi_scf_pinned}), and (\ref{eq:u_scf_general}) make up the system of self-consistent field equations which is solved iteratively: one assumes an initial volume fraction profile $\varphi(\mathbf{r})$, then computes the potential $u(\mathbf{r})$ using Eq.~(\ref{eq:u_scf_general}), the set of the Green's functions, Eq.~(\ref{eq:EDE}), and derives a new volume fraction profile $\varphi\,'(\mathbf{r})$, Eq.~(\ref{eq:phi_scf_pinned}). The procedure is then repeated until the sequence of approximations $\varphi(\mathbf{r}) \to \varphi\,'(\mathbf{r}) \to \ldots$ converges to a stable solution $\varphi\,^*(\mathbf{r})=\varphi(\mathbf{r})$.

To solve the self-consistent field equations rigorously, it is necessary to introduce a numerical algorithm. 
Such numerical scheme invariably involves space discretization (i.e., the use of a lattice). 
Here we follow the method of Scheutjens and Fleer (SF-SCF)~\cite{Fleer:1993}, who used the segment size $a$ as the lattice cell size. 
The lattice sites are organized in \emph{layers}, each of these layers is referred to with a single coordinate $\mathbf{r}$. 
Within a layer, a mean-field approximation is applied, i.e., the volume fractions of the monomeric components and the self-consistent
potential within the layer are constant. 
The way the sites are organized in layers depends on the symmetry in the system and must be preassumed. 
The approach allows for volume fraction and self-consistent field gradients between these layers. 

\begin{figure}[t] 
  \begin{center}
  \includegraphics[width=7cm]{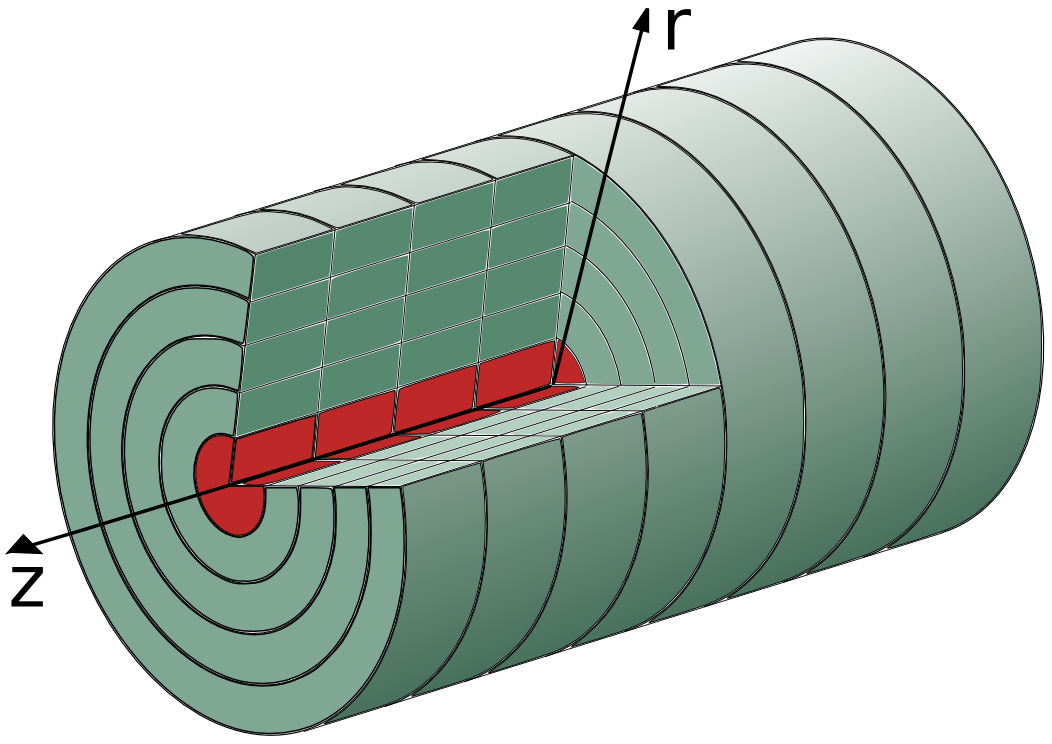}
  \end{center}
  \caption{Cylindrical lattice used in two-gradient SCF calculations.}
  \label{fig:cylbox}
\end{figure}

In order to consider the stretching of a single polymer chain (globule),  it is necessary to use a two-gradient version of the SCF algorithm~\cite{Feuz:2005}, taking into account the symmetry of the problem. The natural geometry for this is a cylindrical coordinate system for which $\mathbf{r}=(r, z)$, Figure~\ref{fig:cylbox}. In this case, all volume fraction profiles as well as other thermodynamic values depend explicitly on the radial coordinate $r$ and the axial coordinate $z$. The system is rotationally invariant with respect to the $z$-axis and the mean-field approximation is applied along the angular coordinate. Therefore a lattice layer $\mathbf{r}=(r,z)$ represents a piece of a tube (a ring) of thickness and height both equal to the lattice unit length $a$.

A polymer chain is represented on the lattice as a freely-jointed chain walk. 
The law of such walk on the cylindrical lattice is specified by setting the \emph{a priori} transition probabilities, 
which are the statistical weights of the steps from the layer with the coordinate $(r, z)$ to a neighboring one $(r+i\cdot a, z+j\cdot a)$, 
where $i,j\in\{-1,0,1\}$, denoted by  $\lambda_{ij}$. 
These statistical weights account for possibilities of going to the nearest neighbor (if $i = 0$ but $j\neq 0$, or  $i\neq 0$ but $j = 0$), next to nearest neighbor (if $i \neq 0$ and $j \neq 0$) or staying in the same layer (if $i=j=0$). Since the number of sites in the layer depends on $r$, $\lambda_{ij}$ is $r$-dependent but has no $z$ dependence. The transition probabilities $\lambda_{ij}(r)$ obey the internal balance equation $L(r) \lambda_{1,j}(r) = L(r+a) \lambda_{-1,j}(r+a)$ where $L(r)=\pi(2r -a)/a$ is the number of sites in the layer with the radial coordinate $r$. The probability to go from the layer with radial coordinate $r$ to that with $r+1$ should be proportional to the contact area between these layers $\lambda_{1,j}(r) \sim A(r)$, the latter is given by $A(r)=2\pi r a$. Similarly, $\lambda_{-1,j}(r) \sim A(r-a)$. Using this area we may write
\begin{equation} \label{eq:lambdas_cyl}
  \begin{split}
    & \lambda_{1,0}(r) = \lambda_1 \frac{A(r)}{a^2L(r)}, \quad \lambda_{-1,0}(r) = \lambda_1 \frac{A(r-a)}{a^2L(r)}, \\
    & \lambda_{1,\pm 1}(r) = \lambda_2 \frac{A(r)}{a^2L(r)}, \quad \lambda_{-1,\pm 1}(r) = \lambda_2 \frac{a^2A(r-a)}{L(r)}
  \end{split}
\end{equation}
For the transition probabilities in $z$ direction $\lambda_{0,\pm 1}(r) = \lambda_1 $, the probability to stay in the same ring $\lambda_{0,0}(r) = \lambda_0$. The probabilities $\lambda_{ij} (r)$  should obey the normalization condition
\begin{equation} \label{eq:lambdas_norm}
  \sum_{i = -1,0,1} \sum_{j = -1,0,1} \lambda_{ij}(r) = 1 .
\end{equation}
In particular, this gives for the ``limiting'' transition probabilities $\lambda_0 + 4 \lambda_1 + 4 \lambda_2 = 1$ (at zero curvature). 
The set of $\lambda_0$, $\lambda_1$, and $\lambda_2$ determines the character of the walk on the lattice,i.e., the chain entropy and rigidity.  
For instance, setting $\lambda_2=0$ eliminates next-to nearest neighbor steps.

In the lattice approach, the iterative procedure of solving the system of the SCF equations is implemented as follows. Once the initial guess for the volume fraction profile, $\varphi(\mathbf{r})$, is set, the self-consistent segment potential $u(\mathbf{r})$ is calculated as follows
\begin{equation} \label{eq:u_potetntial}
  \frac{u(\mathbf{r})}{k_B T} = \log (1 - \varphi(\mathbf{r})) - 2\chi \left\langle \varphi(\mathbf{r}) \right\rangle .
\end{equation}
where $\chi$ is the Flory-Huggins parameter describing the polymer-solvent interaction. The angular brackets in (\ref{eq:u_potetntial}) denote a local \emph{layer average} over the nearest and next-to-nearest lattice layers
\begin{equation} \label{eq:site_average}
  \left\langle X(\mathbf{r})\right\rangle = \left\langle X(r, z)\right\rangle = 
  \sum_{i = -1,0,1} \sum_{j = -1,0,1} \lambda_{ij}(r) X(r+i\cdot a, z+j\cdot a).
\end{equation}
Here the set of transition probabilities $\{ \lambda_{i,j}(r) \}$ 
introduced above for specifying the lattice walk is also used as the set of weight coefficients at averagung.

The Green's functions $G(\mathbf{r}', \mathbf{r};  n)$ can be computed from the recurrence relation expressing the fact that a chain of $n$ monomers can be obtained by adding a monomer to a chain of $n-1$ monomers:
\begin{equation} \label{eq:G_recurrence}
  G(\mathbf{r}', \mathbf{r};  n) = \left\langle G(\mathbf{r}', \mathbf{r}; n-1)\right\rangle G(\mathbf{r}; 1) ,
\end{equation}
where $G(\mathbf{r}; 1)$ is the partition function of a monomer which is simply given by the 
Boltzman law: $G(\mathbf{r}; 1) = \exp [-u(\mathbf{r})/k_BT]$. 
The angular brackets denote here the weighted sum over the neighbours of the layer $\mathbf{r}=(r, z)$ 
which is calculated alike the nearest and next to nearest neighbor average, Eq.~(\ref{eq:site_average}).

Note that the set of $\{ \lambda_{i,j}(r) \}$ (or, equivalently, $\{ \lambda_0, \lambda_1, \lambda_{2} \}$) may be in principle different for density average and the Green's function calculation. Indeed, one might want to exclude the ``diagonal'' or next-to-nearest neighbor steps and set $\lambda_2$ to zero for the \emph{walk} (i.e. in the calculation of $G(\mathbf{r}', \mathbf{r};  n)$), but take into account next-to-nearest neighbour \emph{interactions} and set the corresponding $\lambda_2$ nonzero. 

\subsection{Implementation of the SCF approach} \label{sec:SCF:implement}

Numerical calculations using the SF-SCF approach described above were implemented using \texttt{sfbox} software developed in Wageningen University~\cite{vanMale:2003}. It allows to perform efficient high-speed calculations, even on a personal computer. \texttt{sfbox} uses the same set of $\{ \lambda_0, \lambda_1, \lambda_{2} \}$ both for modelling the chain walk, Eq.~(\ref{eq:G_recurrence}), and performing the neighbor average, Eq.~(\ref{eq:site_average}). 
For our calculations $\lambda_0 = \lambda_1 = \lambda_2 = 1/9$ were chosen. This means that both nearest neighbor and next-to-nearest neighbor layers are taking into account in performing site average, Eq.~(\ref{eq:site_average}) as well as in  modeling the chain trajectory (calculating Green's function), Eq.~(\ref{eq:G_recurrence}).

In the calculations, a cylindrical lattice with limited size is used, therefore a simulation box, 
i.e. the range for $r$ and $z$: $r\in [a, r_{max}]$ and $z\in [a, z_{max}]$,  as well as the boundary conditions should be properly set. The simulation box should not be too large to make the calculations low time- and memory consuming. On the other hand, it cannot be too small, in order to avoid the edge-effects. In all cases we used $r_{max}$ larger then the radius of the unperturbed globule $R_0$. The system size in the $z$-direction depends on the value of the extension. 

The macromolecule is placed symmetrically in the box so that its ends are pinned at the $z$-axis, $\mathbf{r}_1=(a, z_1)$, $\mathbf{r}_N=(a, z_N)$, equidistant from the corresponding boundaries: $z_1 - a = z_{max}-z_N > 2R_0$. The distance between the chain ends is $D=z_N - z_1$.

We followed a specific protocol to study the unfolding of the globule. The first calculation (the first run of \texttt{sfbox}) is made for the minimum distance between the ends of the chain $D/a=1$. Then both $z_N/a$ and, correspondingly, the size of the box $z_{max}/a$ direction are successively increased by unity and the following run is made, etc. The solution obtained at $i$-th step is used as initial guess for $(i+1)$-th step. This procedure is repeated up to strong stretching, $D/a \sim N$ (in practice - up to $D/a \approx N/2$). 

Then a second series of runs is made. This series corresponding to the \emph{refolding} of the globule. It starts from $D/a \sim N$. Subsequently both $z_N/a$ and $z_{max}/a$ are successively decreased by unity down to $D/a = 1$. Again, the solution obtained at the previous step is used as initial guess in the following run. For each run we obtain the free energy and detailed profiles of the volume fractions.

As a result, two free energy dependences on the chain extension and two sets of teh force-deformation profiles, i.e. for the forward (unfolding) 
and backward (refolding) runs, respectively, are obtained for each given pair $\{N, \chi \}$. Note that in the described scheme of SF-SCF modeling only one chain end corresponding to the $N$-th monomer is moved, while the first monomer remains pinned in the same point ($z_1$ remains unchanged). 

There are the following parameters in our model. The number of monomers units is equal to $N$. 
Unoccupied lattice sites are taken by a monomeric solvent (incompressible system) and the corresponding Flory-Huggins parameters for the polymer-solvent interactions is $\chi$.

\section{Results} \label{sec:results}
We have performed SCF calculations for several chain length $N$ ranging from 200 to 1000 and a series of $\chi$ values ranging from $\chi=0.8$ to $2.0$. In the SCF calculations, the results of two types are obtained: (1) macroscopic (large-scale) thermodynamic properties, first and foremost the free energy, and (2) polymer and solvent volume fraction distribution profile (i.e. local properties).

First we consider how the free energy of the globule behaves with a consecutive increase or decrease in the distance $D$ between 
the ends of the chain (i.e. upon the mechanical unfolding or refolding, respectively, of the globule) 
and obtain the force-extension curves - force vs. deformation dependences - for different $N$ and $\chi$ values.  
Then, by considering the evolution of the density profiles, we analyze what conformational changes occur in the globule upon deformation and correlate these with the features of the force-extension curve. 

\subsection{Free energy curves} \label{sec:results:FE}
\begin{figure}[t] 
  \begin{center}
  \includegraphics[width=8cm]{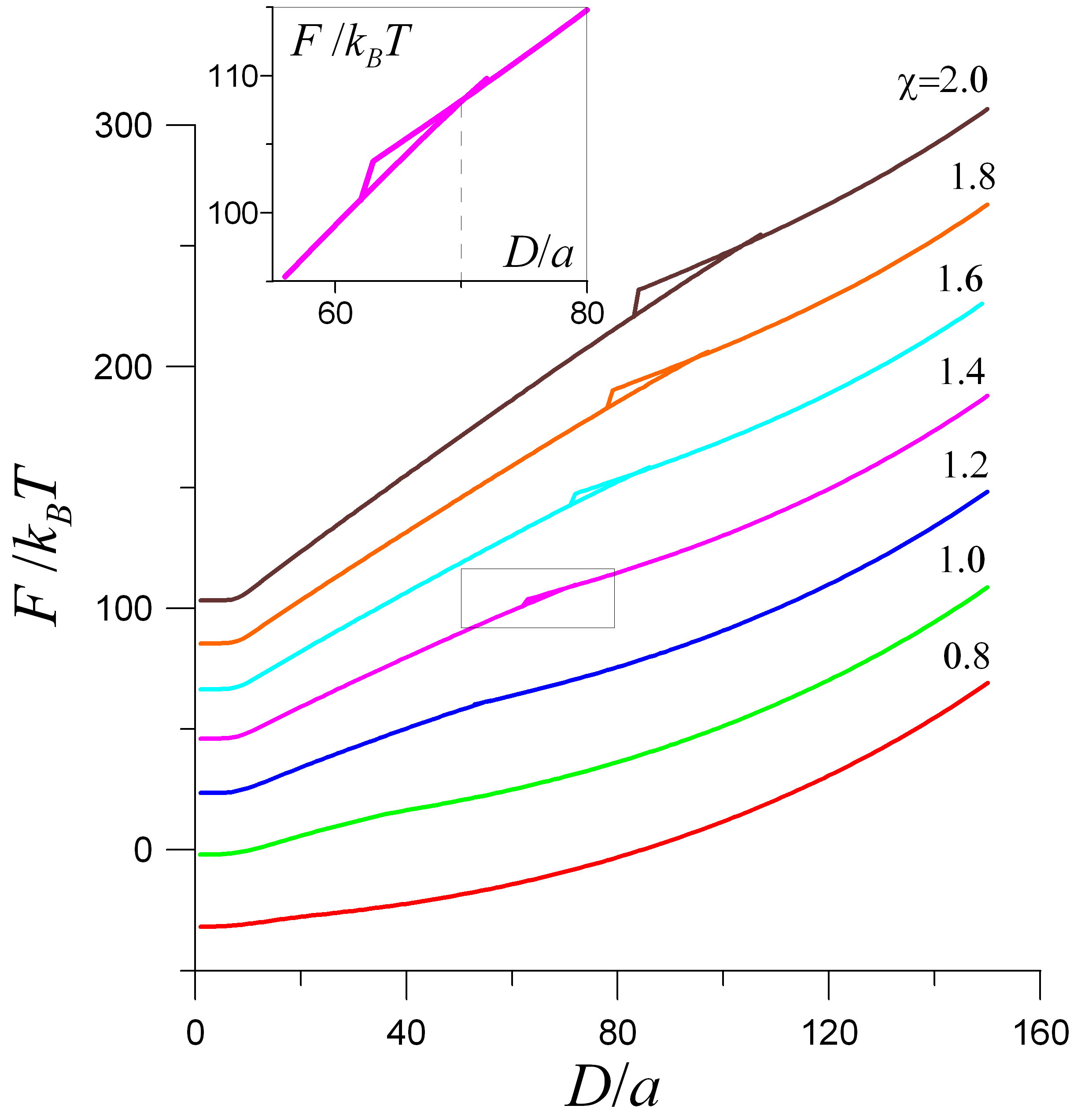}
  \end{center}
  \caption{Free energy as function of chain extension for $N=200$ and various values of $\chi$. 
           Inset: free energy for $\chi=1.4$ in the vicinity of the transition point.}
  \label{fig:FE_200}
\end{figure}

Figure~\ref{fig:FE_200} shows an example of the free energy $F$ as a function of the extension $D$ calculated for $N=200$ and a series of $\chi$-values. 
Both unfolding and refolding branches are plotted in this figure and, as one can see, 
two branches coincide both at small and high stretching values $D$, whereas at moderate extensions 
they differ within some range of $D$-values provided the Flory parameter $\chi$ is high enough. 
This means that at moderate deformations, there exist two local minima of the free energy. 
The position and the width of the region where these two minima coexist depends on the $\chi$-value. 
The free energy is a functional of the density distribution $F=F[\varphi (r,z)]$ and the two minima correspond to different states of the deformed globule, and there are two different (locally) equilibrium volume fraction profiles $\varphi_1(r,z)$ and $\varphi_2(r,z)$. The state that has a lower free energy is stable (global minimum), the other state is metastable. At the point where the two minima have the same depth, which occurs when the unfolding and the refolding branches intersect, $D=D_{tr}$, the system suffers (on the mean-field level) a first-order-like phase transition. 

However, the two minima are separated by a free energy barrier. 
For finite chain length this barrier has a finite height and in reality the transition is smooth; 
the system can fluctuate between local minima. However in SCF calculations, 
where thermal fluctuations are suppressed, the system stays in the metastable state until it 
reaches the \emph{spinodal point}, where the metastable minimum disappear, and then jumps to the other (global) minimum. 
Hence, the mechanical unfolding-refolding cycle exhibits in a SCF calculation a hysteresis loop.

At lower values of $\chi$ ($\chi=0.8$ and $1.0$) the dependence of the free energy on deformation is different: 
unfolding and refolding branches of the free energy superimpose completely.

\subsection{Force-extension curves} \label{sec:results:FEC}
\begin{figure}[t] 
  \begin{center}
  \includegraphics[width=8cm]{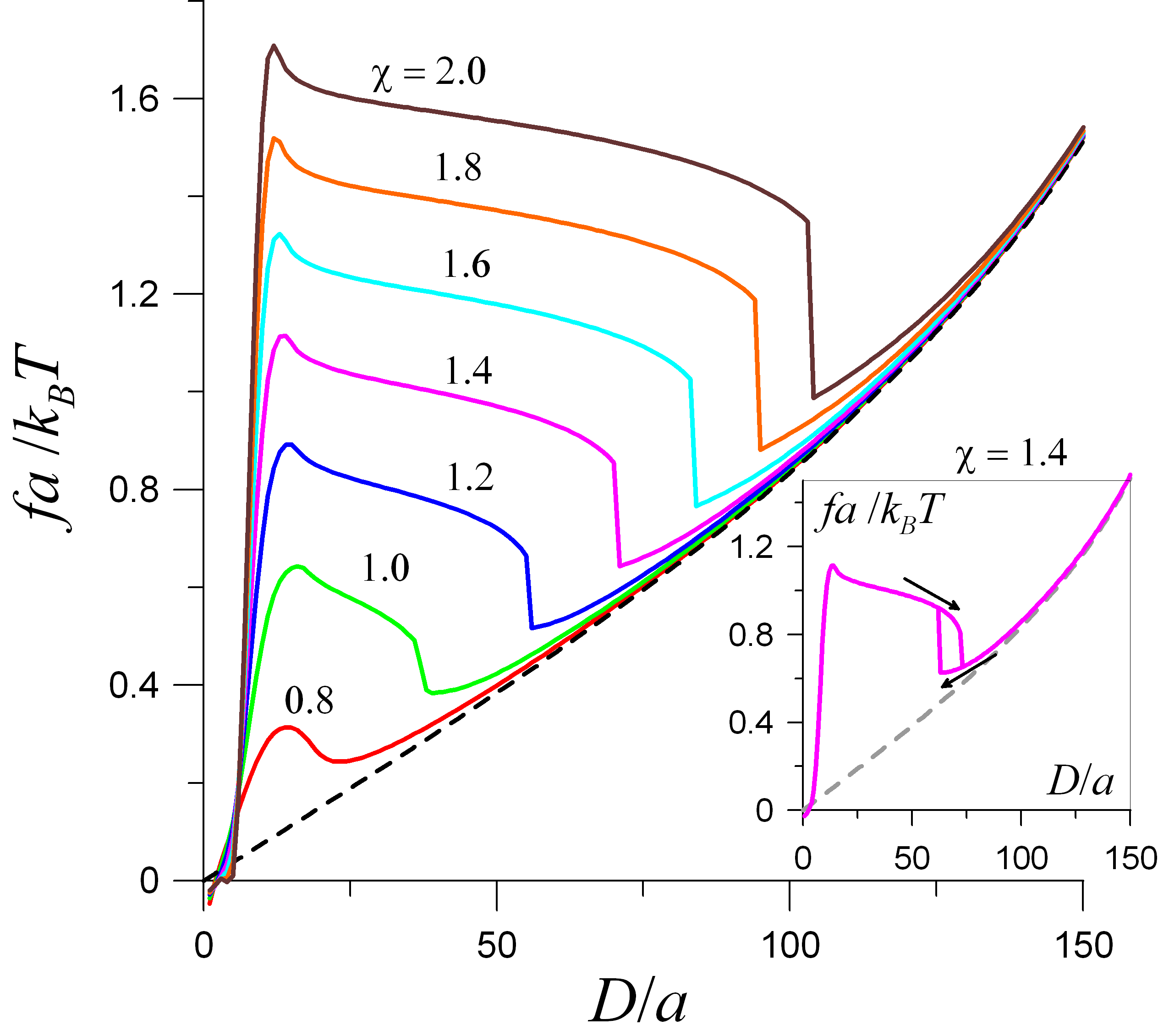}
  \end{center}
  \caption{Equlibrium reaction force vs. extension curves for the globule with $N=200$ at various values of $\chi$. Inset: Reaction force calculated for forward (unfolding) and backward (refolding) runs for $\chi=1.4$.}
  \label{fig:force_200}
\end{figure}

Once the dependence of the free energy vs. extension is known, 
the reaction, or restoring, force can be found by differentiating the free energy $F$ 
with respect to end-to-end distance $D$. The value of the reaction force is $f=\partial F/\partial D$ and it acts against the extension. 
Figure~\ref{fig:force_200} shows force-extension curves obtained from the free energy dependences on deformation, 
Figure~\ref{fig:FE_200}, by numerical differentiation of the \emph{equilibrium} free energy. 
The true transition point (in thermodynamic sense) $D_{tr}$ is obtained from the condition $F_1(D_{tr}) = F_2(D_{tr})$ 
whereas the metastable states are excluded from these consideration. 
The kink in the free energy curves at the transition point gives rise to a jump in the reaction force.

If the metastable states are taken into account, 
i.e. unfolding and refolding free energy branches are different, 
one obtains the hysteresis loop on the force-extension curve. As an example one of such curves is 
shown for $\chi=1.4$ in the inset of Figure~\ref{fig:force_200}.

Inspection of the force-extension curves clearly reveals three different deformation regimes. 
(1)~At small deformations the reaction force strongly increases with extension. Note that at very small extension, $D/a \simeq 1$, the force is negative: Fixing the chain ends to very small distances imposes high local concentration of the monomer units that is larger than the average equilibrium concentration in the globule; this is penalized by an increase in the free energy and this causes a repulsion between the end-monomers. 
(2)~After reaching a peak value the reaction force slightly decreases (quasi-plateau) in a wide range of $D/a$ 
and then drops down at the transition point $D=D_{tr}$, where the equilibrium free energy dependence has a kink, 
Figure~\ref{fig:FE_200}. Neither the extensive plateau nor the jump in the force are observed in the case of $\chi=0.8$. 
This is consistent with observation that the unfolding and 
refolding branches are equivalent and that there was no kink~/~transition point in this case. 
Interestingly, at $\chi=1.0$ the unfolding and refolding free energy branches coincide  too, 
but the corresponding force-extension curve shows a pronounced jump. 
This can be attributed to the narrow width of the region where the two free energy minima coexist;  
this region cannot be resolved within our lattice approach since the latter has 
the resolution $\Delta D/a \geq 1$. (3)~After the jump, at strong deformations, the reaction force starts to grow again. 
The force-extension dependence in this regime is universal and is independent of the solvent quality $\chi$. 
This free energy and hence the reaction force have a purely entropic origin. 
As is discussed in Appendix, for a lattice chain, an analytical expression for the force dependence on deformation exists 
for this branch (in the whole range of chain extensions). This analytical result is shown in Figure~\ref{fig:force_200} by the dashed line.

At this stage we mention that the transition between regimes (1) and (2) occurs continuously. 
In summary, an increase in $\chi$ leads to (i) an increase of the reaction force at fixed end-to-end distance; 
(ii) the broadening of the quasi-plateau (regime (2) ), and (iii) an  increase in the value of the jump of the force at the transition point.

\subsection{Density profiles} \label{sec:results:profiles}
\begin{figure}
  \begin{center}
    \includegraphics[width=12cm]{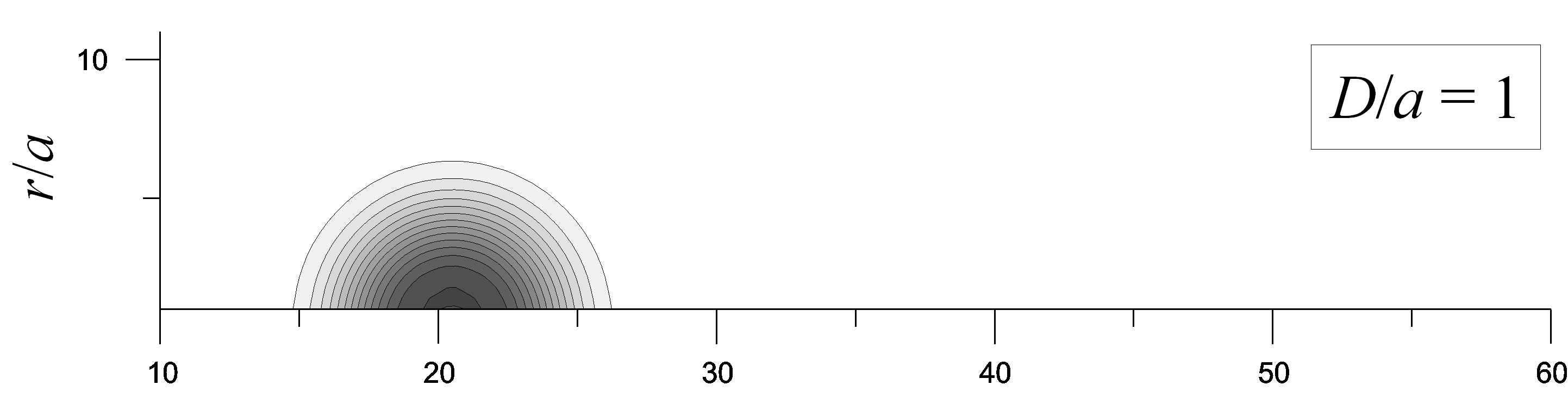}
    \includegraphics[width=12cm]{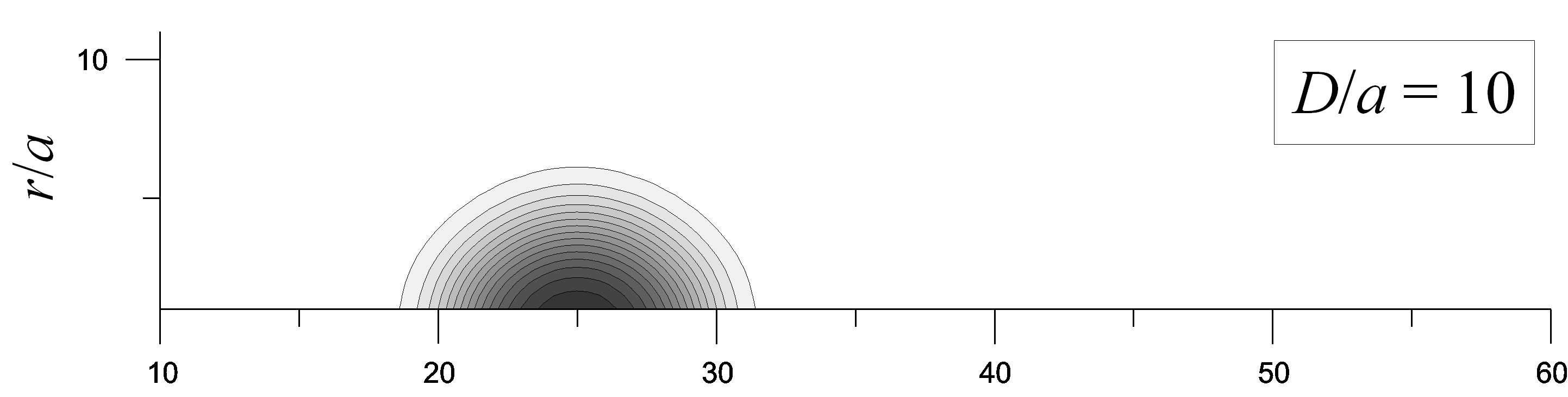}
    \includegraphics[width=12cm]{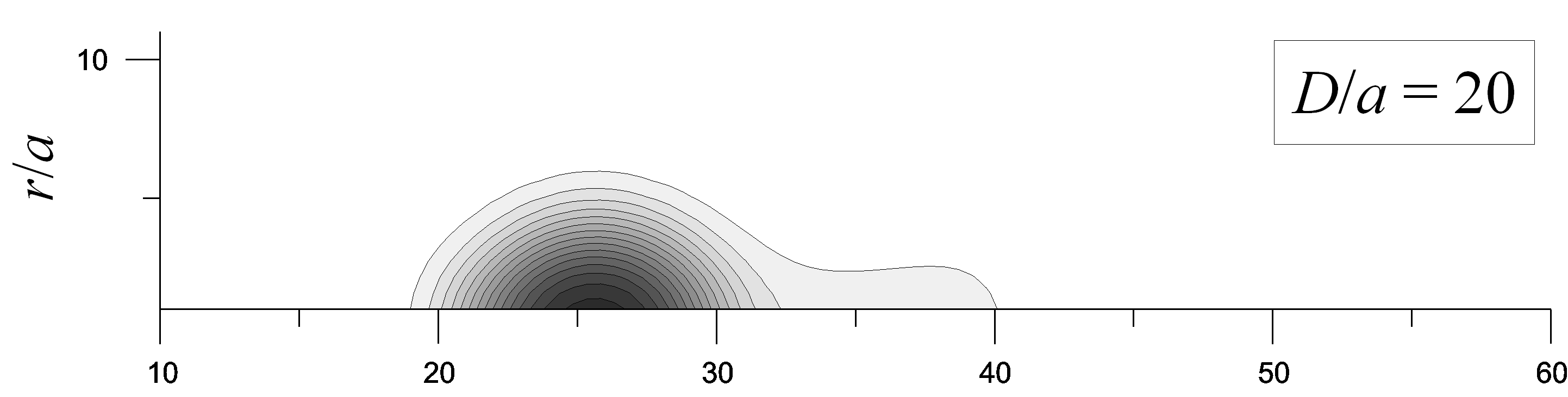}
    \includegraphics[width=12cm]{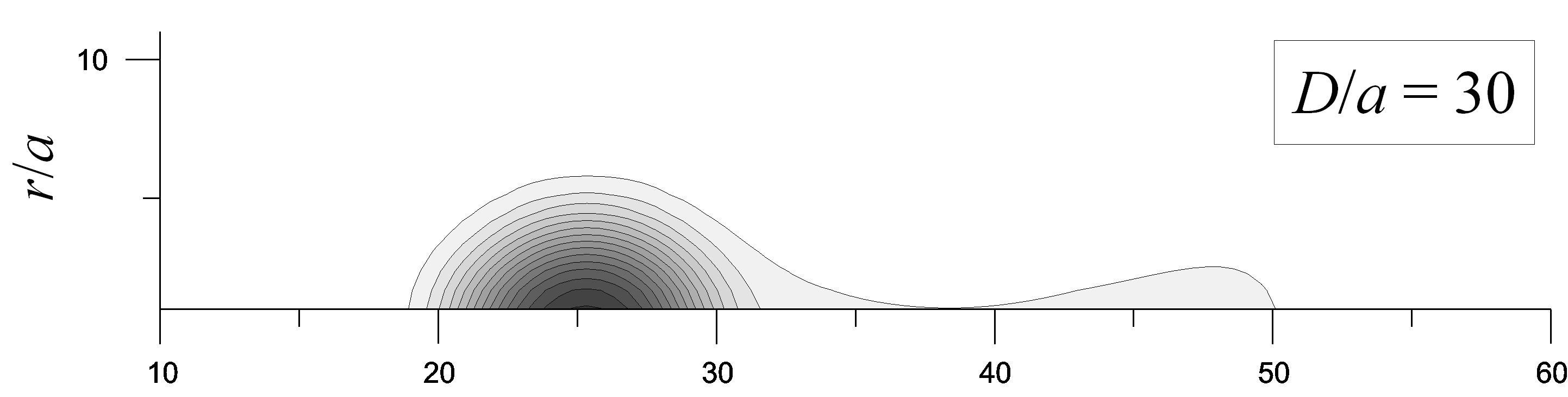}
    \includegraphics[width=12cm]{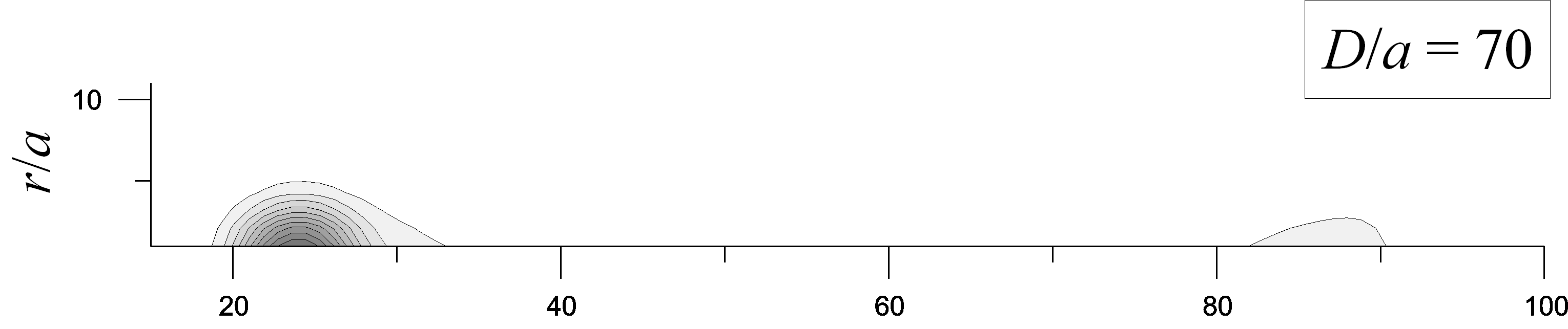}
    \includegraphics[width=12cm]{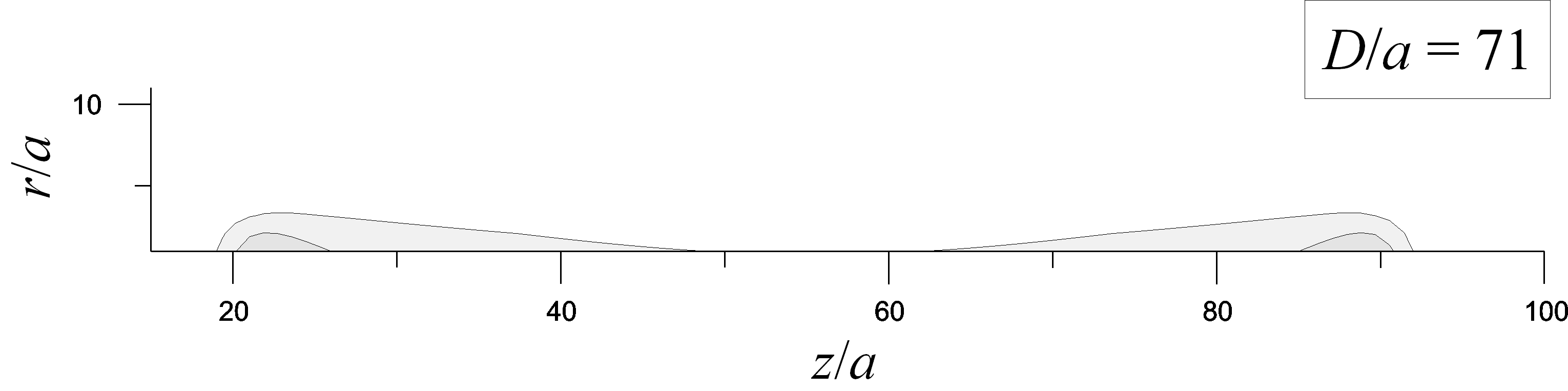}
    
    \vspace{1cm}
    \includegraphics[width=8cm]{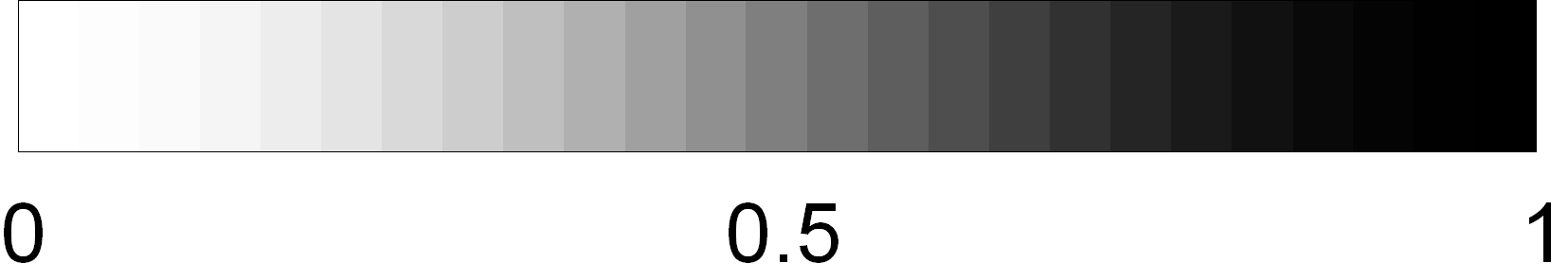}
  \end{center}
  \caption{Polymer density profiles $\varphi(r,z)$ for the globule with $N=200$ and $\chi=1.4$ at various chain extensions.}
  \label{fig:profiles_200_chi14}
\end{figure}

To correlate the observed deformation regimes with (possible) conformational changes in the globule, 
profiles of polymer volume fraction distribution in the deformed globule should be analyzed. 
As it was mentioned above, the SF-SCF approach gives access to these polymer density distribution profiles $\varphi (r,z)$. 
The volume fraction is a function of two variables that can best be presented in two-dimensional contour plot, Figure~\ref{fig:profiles_200_chi14}.

\begin{figure}[t] 
  \begin{center}
    \includegraphics[width=8cm]{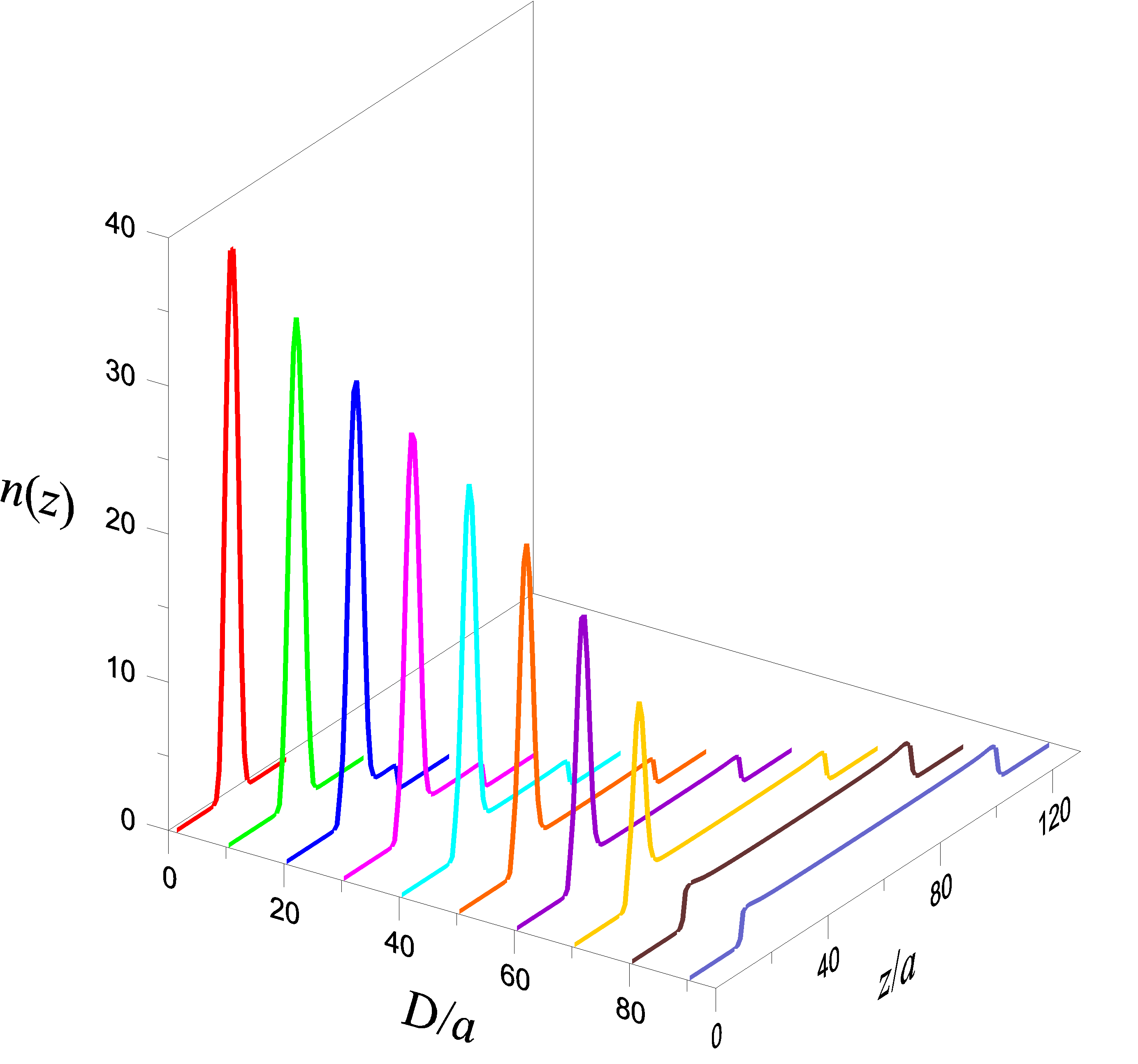}
  \end{center}
  \caption{Number of monomer units per $z$ axis unit length, $n(z)$, for the globule with $N=200$ and $\chi=1.4$ 
           at various chain extensions.}
  \label{fig:nprofiles_200_chi14}
\end{figure}

Another interesting and illustrative quantity that characterizes the conformations of the deformed globule is the 
``integral'' profile - the axial distribution of monomer units (the number of monomer units per $z$ axis unit length)
\begin{equation} \label{eq:nz}
  n(z)=\sum_{r\geq a} L(r)\varphi(r, z) .
\end{equation}
A set of $n(z)$ profiles for different extensions can be conveniently represented as a 3-dimensional plot,  Figure~\ref{fig:nprofiles_200_chi14}.

Figures~\ref{fig:profiles_200_chi14} and \ref{fig:nprofiles_200_chi14} show the evolution of 
$\varphi (r,z)$ and $n(z)$ upon the imposed deformation for $\chi=1.4$ and $N=200$. 
This choice of $\chi$ and $N$ represents the typical case where three regimes of globule deformation, Figure~\ref{fig:force_200}, are well distinguished. 
One can see that at small extensions (regime (1)) the globule changes its shape from spherically symmetric (at $D/a=1$), 
to an asymmetric one similar to that of a prolate ellipsoid (at $D/a=10$). 
When the extension grows, the prolate globule conformation becomes unstable and the globule splits into a dense ``head'' and a stretched ``tail'' 
coexisting in one macromolecule (at $D/a=20$), thus acquiring a tadpole conformation. 
This corresponds to regime (2) on the force-extension curve. 
When the distance between the ends of the chain increases, a redistribution of monomers between two phases occurs: 
the tail length grows and the head size decreases, this corresponds to the range of extensions from approximately $D/a=20$ to $D/a=70$. 
In spite of the decrease in the size of the globular head, the density of the globular core (i.e., except of the density in the diffuse interfacial layer) 
remains virtually constant. Close inspection reveals that the number of monomer units per unit length in the tail weakly increases. 
This is in accordance with a very weak decay of the reaction force $f$ in the plateau regime, Figure~\ref{fig:force_200}, because the number of monomers per unit length is the inverse of chain extension: $f\sim dz/dn = 1/n(z)$.

At a certain extension (at $D/a\approx 71=D_{tr}/a$, in the transition point) the globular head disappears. 
Now the chain gets completely unfolded and one enters regime (3) (Figure~\ref{fig:profiles_200_chi14} at $D/a=71$). 
The disappearance of the globular head in the transition point leads to 
(i) a gain in the surface energy (since the interface disappears), 
(ii) a gain in the conformational entropy (since the chain tension decreases) and 
(iii) a penalty in the volume interaction free energy (since the monomer units that constituted the globular head get exposed to 
the solvent). The mutual cancellation of these three contribution determines the transition point. 
At larger extensions the tadpole conformation is metastable and the globule completely unfolds. The size of the disappearing globule in the transition point is still quite large (at $D/a=71$ the numbers of monomer units in the tail and in the head can be estimated from Figures~\ref{fig:profiles_200_chi14} and \ref{fig:nprofiles_200_chi14} as $n_{tail}\approx 120$, $n_{head}\approx 80$).

The density profile in the strong stretching regime has the shape of a homogeneous sphero-cylinder 
with local maxima in the points where the chain end are fixed.

\subsection{Special case: $\chi=0.8$, $N=200$ } \label{sec:results:special}

The case $\chi=0.8$, $N=200$ calls for special attention. 
The force extension curve, Figure~\ref{fig:force_200}, has an untypical shape compared to those for larger $\chi$ values: 
there is no quasi-plateau regime at intermediate extensions and the force jump is missing. 
In this case the unfolding and refolding branches of the free energy vs. deformation coincide, see Figure~\ref{fig:FE_200}.

\begin{figure}
  \begin{center}
    \includegraphics[width=12cm]{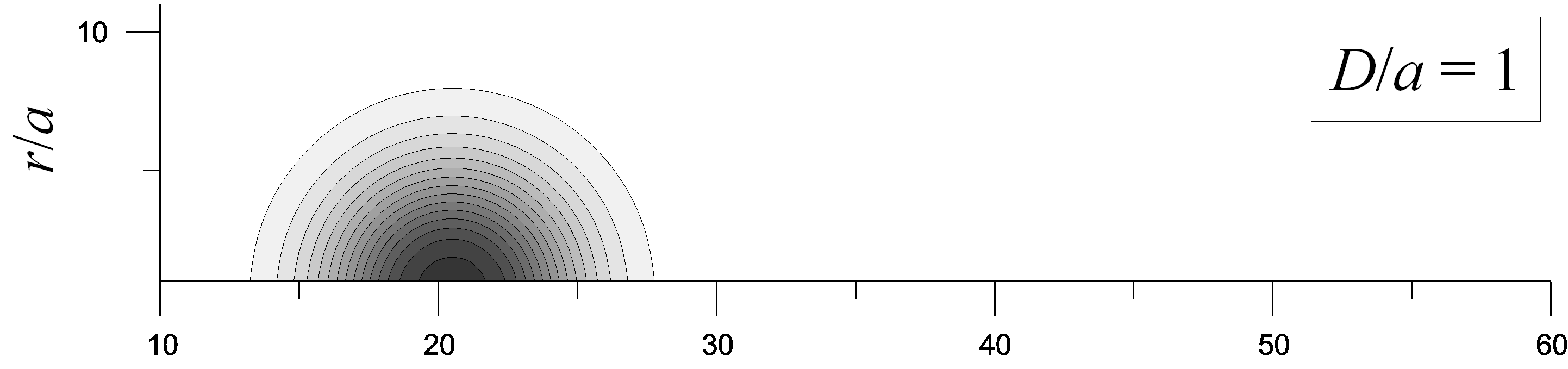}
    \includegraphics[width=12cm]{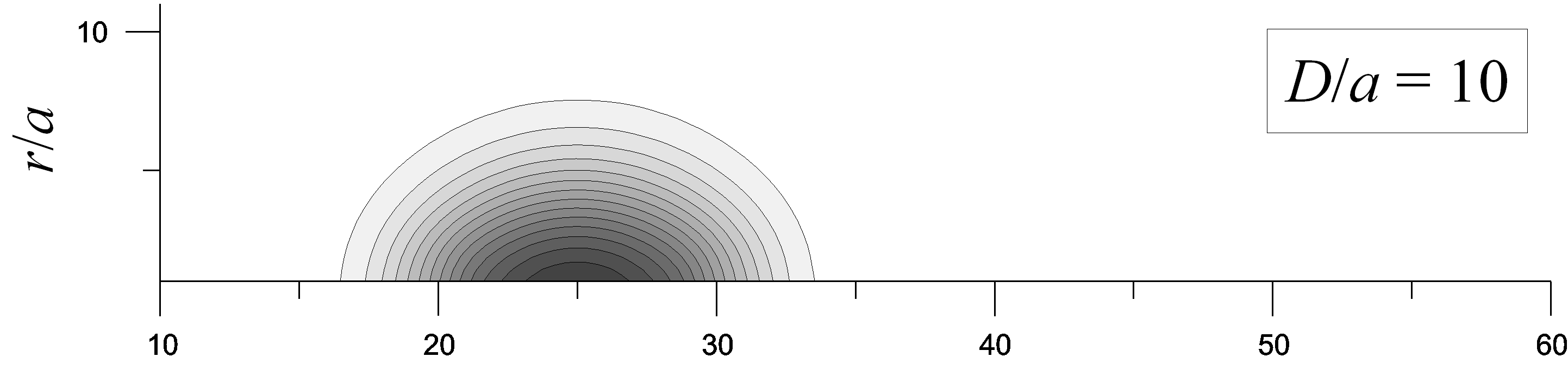}
    \includegraphics[width=12cm]{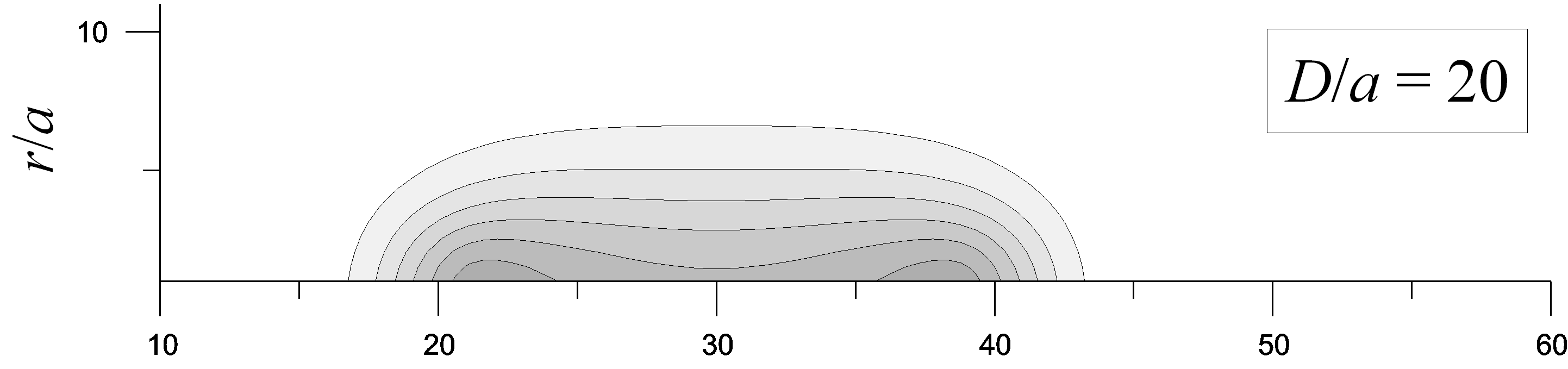}
    \includegraphics[width=12cm]{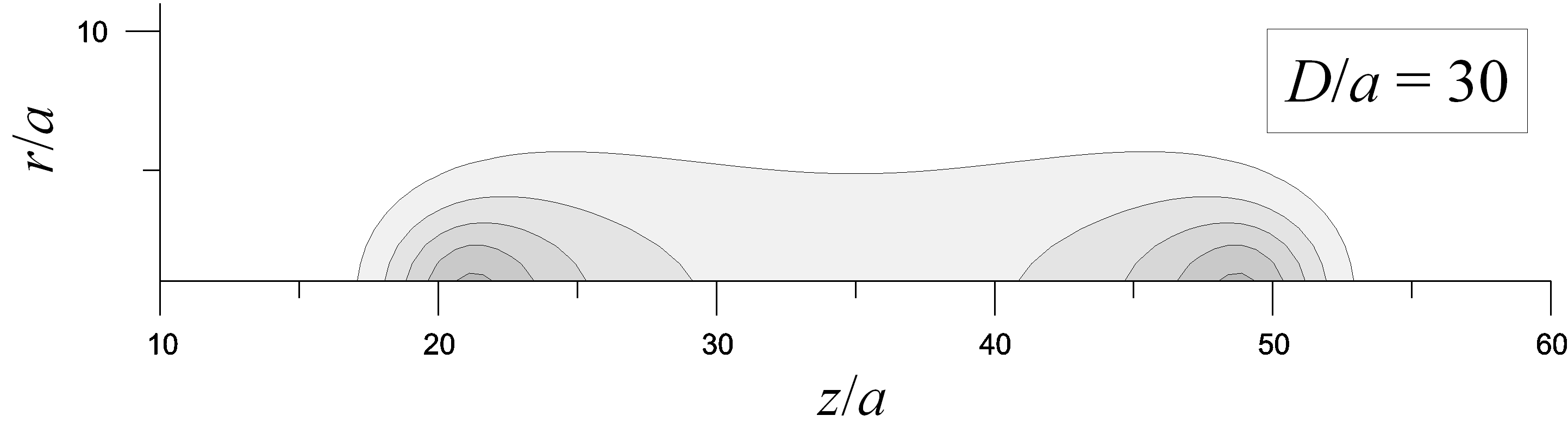}
    
    \vspace{1cm}
    \includegraphics[width=8cm]{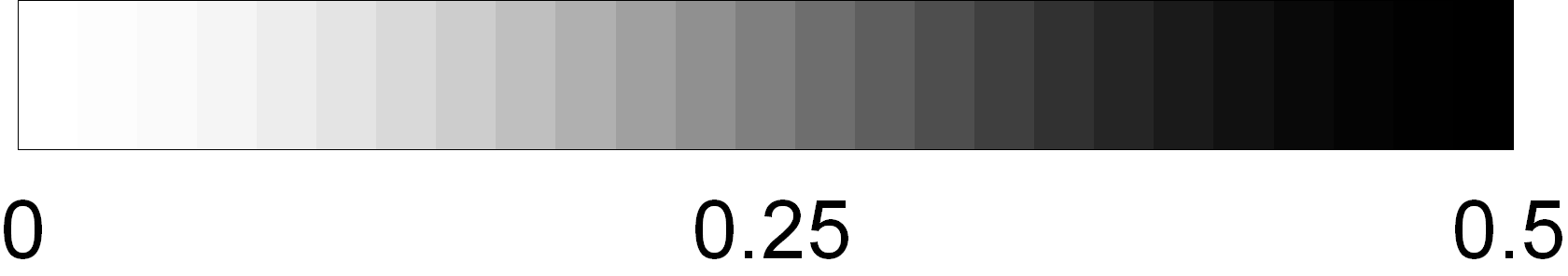}
  \end{center}
  \caption{Polymer density profiles $\varphi(r,z)$ for the globule with $N=200$ and $\chi=0.8$ at various chain extensions.}
  \label{fig:profiles_200_chi08}
\end{figure}

\begin{figure}[t] 
  \begin{center}
    \includegraphics[width=8cm]{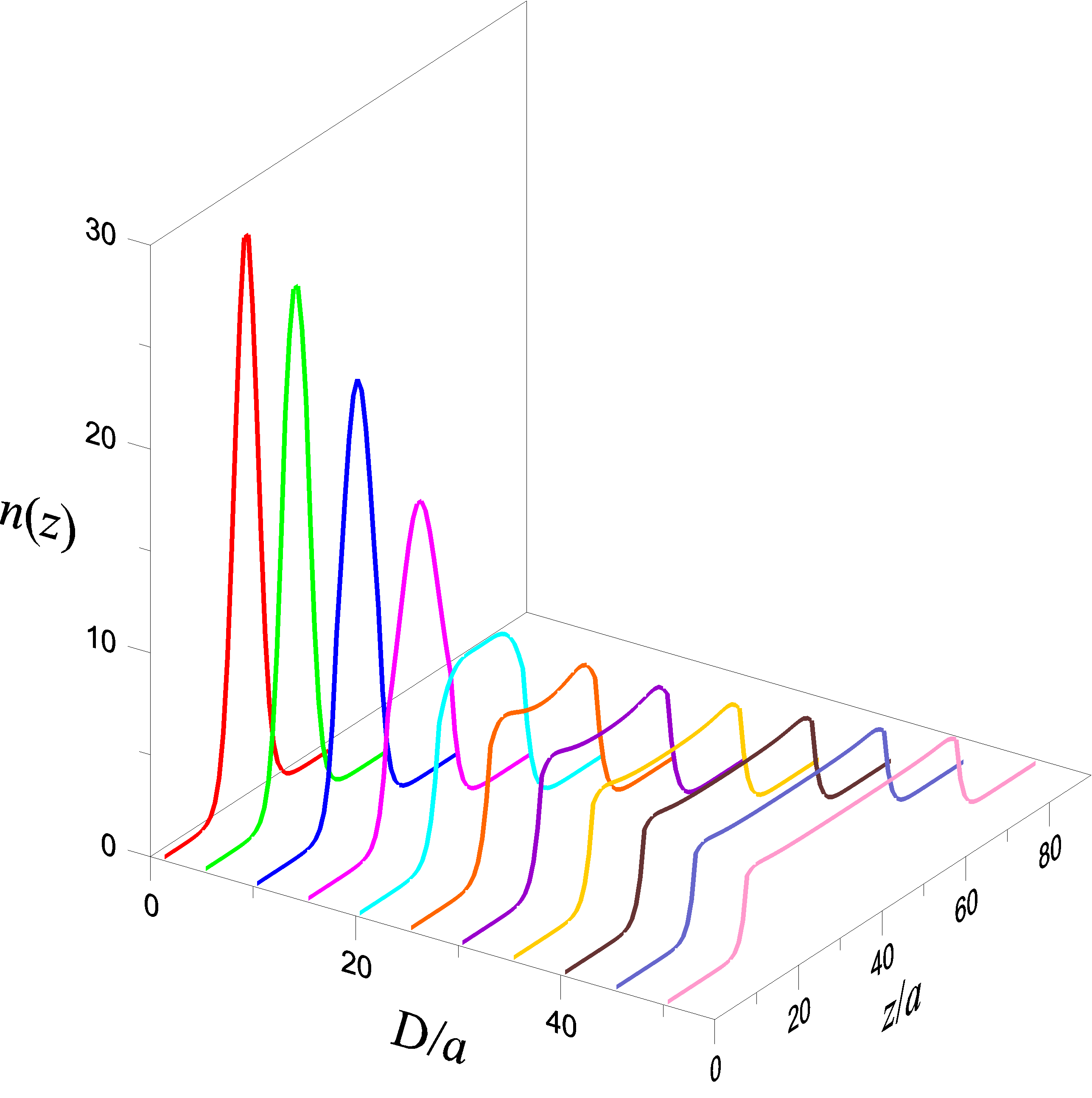}
  \end{center}
  \caption{Number of monomer units per $z$ axis unit length, $n(z)$, for the globule with $N=200$ and $\chi=0.8$ 
           at various chain extensions.}
  \label{fig:nprofiles_200_chi08}
\end{figure}

The analysis of the density profiles and the axial monomer distributions, Figures~\ref{fig:profiles_200_chi08} and \ref{fig:nprofiles_200_chi08}, 
respectively, demonstrates that in the considered case we encounter an essentially different unfolding mechanism: 
upon stretching the globule becomes more asymmetric and the density of its core decreases. What is also important, this transformation occurs \emph{continuously}.

The reason of such a behavior can be explained as follows. 
The globule has an interfacial layer with a finite thickness which increases as $\chi$ decreases (as the solvent becomes better for the polymer).  
In the case of a short chain, the width of the interface becomes comparable to the size of  the globule (radius). 
It appears that it is thermodynamically more preferable to reduce the density in the globular core 
rather than to form an extended tail keeping the (high) core density. 
Therefore, instead of a redistribution of the monomer units between a collapsed (globular) and a stretched phase, these are transferred from the core to the interfacial layer, as $D$ increases.

\subsection{Effect of the polymerization degree on the globule deformation} \label{sec:results:alln}

\begin{figure}[t] 
  \begin{center}
    \includegraphics[width=8cm]{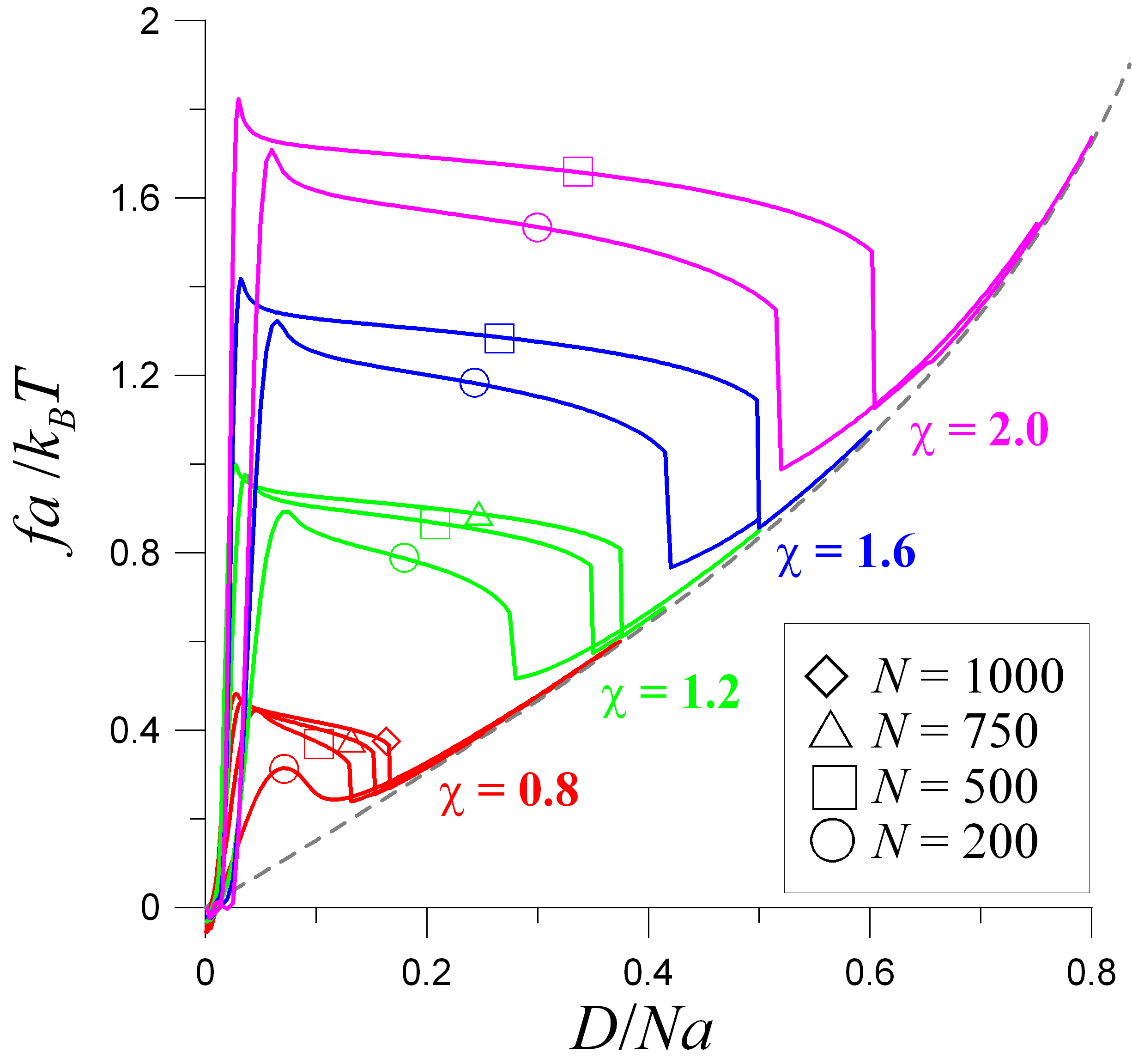}
  \end{center}
  \caption{Force extension curves at various values of $N$ and $\chi$.}
  \label{fig:force_all}
\end{figure}

Let us consider in more detail the influence of the degree of polymerization $N$ and polymer-solvent interaction parameter $\chi$ 
on the force-extension curves. 
Figure~\ref{fig:force_all} presents the force-extension curves calculated for different values of $N$ and $\chi$. 
In order to compare the data for different $N$ the forces were plotted vs the reduced extension $D/(Na)$. 
Figure~\ref{fig:force_all} shows that for all considered cases, except of $\chi=0.8$, $N=200$, three deformation regimes are observed: 
(i) a linear force growth with an increase in $D/(Na)$ at small extensions and (ii) a quasi-plateau in a wide range of $D/(Na)$ values ending by a sharp decrease in $f$ followed by (iii) a universal force-extension dependence of the ideal chain (see Appendix~\ref{app:Fconf}, Eq.~(\ref{eq:Dchain_f})). The position of boundaries between the regimes and the dependences of the restoring force on the deformation in regimes (i) and (ii) are determined by $N$ and $\chi$, the force-extension relation in regime (iii) depends only on $N$. The $\chi$-dependence was already discussed for $N=200$, Figure~\ref{fig:force_200}. 
Analysis of Figure~\ref{fig:force_all} shows that an increase in either $N$ or $\chi$ has a similar qualitative effect on the shape of the 
force-extension curve. With an increase in $\chi$ or $N$ the 
quasi-plateau region broadens mostly due to a noticeable shift of its right-hand boundary to larger $D/(Na)$ 
values and the weak displacement of its left-hand boundary to smaller $D/(Na)$. 
Therefore, it can be concluded that an increase in the chain length favors micro-segregation within the deformed globule. 
In particular we see that, in contrast to the case of $N=200$ considered above, at $\chi=0.8$ the globule formed by a longer chain  
unfolds via the formation of a micro-segregated (tadpole) structure with a subsequent drop in the reaction force. The height and the decaying slope of quasi-plateau also changes:  with an increase in $N$ and/or $\chi$ it shifts up and flattens. Upon a progressive increase in $N$ the plateau level approaches some asymptotic height controlled by the value of $\chi$, whereas for given value of $N$ it monotonously increases as a function of $\chi$.

Taking into account the analysis of the conformations for $N=200$ as discussed above, we can presume that the picture of the unfolding of the globule as predicted by the SF-SCF method 
retains the main general features such as (i) the three stages of unfolding (three deformation regimes), i.e., (cf Figure~\ref{fig:globules}) the growth of the force at small and strong deformations and its weakly decaying ``quasi-plateau'' behavior in the intermediate micro-segregated regime; (ii) a continuous microphase segregation (ellipsoid-tadpole) transition and the drop of the force in the second transition point where the micro-segregated structure becomes unstable and the globule completely unfolds. 
When the microphase segregation takes place, the head of the tadpole has a clear prolate ellipsoidal and not a spherical shape. On the other hand our SCF results clearly indicate that at moderately poor solvent quality the small globule unfolds without intra-molecular microphase segregated state. 

\section{Blob picture of globule deformation} \label{section:blob}
In this section we discuss the above described results obtained using SF-SCF approach in terms of the blob picture of globule deformation.

\subsection{A blob picture of polymer globule} \label{sec:blob:free}
Let us consider again a single polymer chain with a degree of polymerization $N$ immersed into a poor monomeric solvent. The chain is assumed to be intrinsically flexible, that is, the statistical segment length is of the order of a monomer unit length $a$, the latter coincides with the chain thickness. The binary attractive short-range (van der Waals) interactions between the monomer units are described in terms of Flory-Huggins interaction parameter $\chi > 0.5$. It is well known that in a poor solvent the polymer chain collapses into a spherical globule with a density $\varphi$  which is a function of $\chi$. Under moderately poor solvent strength conditions, that is when $(\chi - 0.5) \ll 1$, the monomer unit volume fraction in the globule scales as $\tau \sim (\chi - 0.5)$. The globule can be envisioned as an array of closely packed Gaussian thermal blobs, Figure~\ref{fig:globules}~a. The blob size, $\xi_t \simeq \tau ^{-1}a$ is determined by the correlation length of the density fluctuations inside the globule. The size of an unperturbed globule scales as 
\begin{equation} 
R_{0}\simeq (N/g_t)^{1/3}\xi_t\simeq (N/\tau)^{1/3}a
\end{equation}
The chain fragments within a thermal blob retain Gaussian statistics, the number of monomer units per thermal blob scales as $g_t\simeq (\xi_{t}/a)^2$. 
The free energy of attractive monomer-monomer interaction per blob is of the order of thermal energy $k_B T$. Therefore the free energy of the globule scales (in the main term or in so called "volume approximation") as 
$F_{globule, v}/k_BT\simeq - N/g_t \simeq - Na^2/\xi_t^2 \simeq - N\tau ^2$. 
A lower order correction in $N$ arises due to the excess free energy of the interface between the collapsed globule and the (poor) solvent: the monomer units that are localized close to the interface exhibit in average more unfavorable contacts with the solvent than monomer units in the interior of the globule. In the scaling approximation an excess free energy of the order of $\simeq k_BT$ is attributed to each thermal blob localized at the globule interface. Hence, the excess free energy of the interface scales as $F_{globule,s}/k_BT\simeq (R_{0}/\xi_t)^2 = (N\tau ^2)^{2/3}$. This excess interfacial free energy stabilizes the spherical shape of a free (non-deformed) globule. 

Thus, thermodynamic properties of the globule are determined by the number of blobs (i.e. by the total number of blobs and by the number of surface blobs), while the globule size is controlled by both the number and the size of the blobs which are the functions of $\chi$.

\subsection{Globule deformation} \label{sec:blob:HZh}
Following the scaling approach of Halperin and Zhulina \cite{Halperin:1991:EL} we now consider a deformed polymer globule as these occur when the end-to-end distance $D$ is imposed.  The  extensional deformation $D > 2R_{0}$  of the globule leads (under the constraint of the conservation of its volume) to an increase in the area of the interface, thus giving rise to an additional free energy penalty. The excess surface area of a weakly deformed globule, Figure~\ref{fig:globules}~b,  grows as $\Delta S\simeq (D-2R_{0})^2$ as long as $(D-2R_{0}) \lesssim R_{0}$ and here subscript $"0"$ refers to the unperturbed globule. Hence, at weak deformations the restoring force that develops upon the extension of the globule, grows linearly with deformation, 
\begin{equation} \label{eq:force_el}
f/k_B T = \partial (F_{globule,s}/k_BT)/\partial D\simeq (D-2R_{0})/\xi_{t}^2
\end{equation}
In the opposite limit of strong extensions, Figure~\ref{fig:globules}~d, $D\geq (N/g_t)\xi_t$, the intra-molecular interactions do not affect the elastic response. Such chain obeys Gaussian entropic elasticity,
\begin{equation} 
f/k_BT\simeq D/Na^2
\end{equation}
and can be presented as a string of  Gaussian elastic blobs~\cite{Pincus:1976}  
\begin{equation} 
D\simeq (N/g)\xi
\end{equation}
of the size $\xi \leq \xi_t$. 
We remark that here we do not consider the non-linear elasticity effects that may occur due to the finite chain extensibility of the chains. 
The latter becomes important at limiting extensions, $D\simeq Na$, see Appendix~\ref{app:Fconf}.

As has been noted by Halperin and Zhulina, in the intermediate range of extensions, $2R_{0}\ll D \ll (N/g_t)\xi_t$, the surface area and the excess interfacial free energy of a homogeneously elongated (cylindrical or prolate ellipsoidal) globule scales as $F_{globule,s}\sim D^{1/2}$. This leads to a {\it decrease} of the restoring force as a function of the extension. The non-monotonic behavior may be identified as a van der Waals loop,  suggesting an intra-molecular co-existence of an extended chain fragment (``tail'') with a depleted spherical globular core, Figure~\ref{fig:globules}~c. According to Halperin and Zhulina the unfolding of the globule in the intermediate range of extensions occurs at fairly constant force, 
\begin{equation} \label{f_c}
f/k_BT\simeq 1/\xi_t
\end{equation}
that corresponds to a tail consisting of a string of elastic blobs with a size equal to that of thermal blobs, $\xi_t$. If finite size corrections are considered, the coexistence plateau $f/k_BT\simeq 1/\xi_t \sim D^0$ is replaced by a region where the force $f$ is a weakly decreasing function of the extension $D$. Indeed, consider the globule in the tadpole conformation, Figure~\ref{fig:globules}~c. Let us assume that the head of the tadpole has a spherical shape (that is, in the scaling analysis we can neglect the asphericity of the head) and is composed of $n$ monomers. Then, the radius of the head scales as $R_{globule}\simeq (n/g_t)^{1/3}\xi_t \simeq (na^2\xi_t)^{1/3}$. The free energy of the tadpole accounts both for the contributions of the globule and the tail: $F_{tadpole} = F_{globule} + F_{tail}$ . The free energy of the globule contains volume and surface terms
\begin{equation} \label{eq:tadp:glob}
  \frac{F_{globule}}{k_B T} \simeq - \frac{n}{g_t} + \frac{S_{globule}}{\xi_t^2} \simeq 
  - \frac{na^2}{\xi_t^2}  + \left(\frac{na^2}{\xi^{2}_t}\right)^{2/3}.
\end{equation}

The tail consists of $N-n$ monomers and can be represented as a stretched string of thermal blobs. The tail length is $D-2R_{globule}$. The number of blobs in the tail scales as $(N-n)/g_t \simeq (N-n) (a/\xi_t)^2$. Therefore the tail length can also be expressed as $D-2R_{globule} \simeq (N-n)a^2/\xi_t$. The free energy of the tail comprises an elastic contribution which is proportional to the number of thermal blobs in the tail
\begin{equation} \label{eq:tadp:tail}
  \frac{F_{tail}}{k_B T} \simeq \frac{(D-R_{globule})^2}{(N-n) a^2} \simeq \frac{(N-n) a^2}{\xi_t^2} .
\end{equation}

As a result, the free energy of the tadpole is
\begin{equation} \label{eq:tadp:all}
  \frac{F_{tadpole}}{k_B T} \simeq \left(\frac{na^2}{\xi^{2}_t}\right)^{2/3} + \frac{(N-2n) a^2}{\xi_t^2}.
\end{equation}

The restoring force is calculated as
\begin{equation} \label{eq:tadp:force}
  \frac{f_{tadpole}}{k_B T} = \frac{1}{k_B T} \cdot \frac{d F_{tadpole}/dn}{dD/dn} \simeq 
  \frac{1}{\xi_t}\left[1 - \left(\frac{\xi_t}{a}\right)^{2/3} \frac{1}{n^{1/3}}  \right] \simeq  
  \frac{1}{\xi_t}\left[1 - \left(\frac{g_t}{n}\right)^{1/3} \right]
\end{equation}
and we notice that as $D$ increases, then both $n$ and thus $f$ decrease.

\subsection{Ellipsoid-to-tadpole transition}
The location of the ellipsoid to tadpole transition can be estimated using simple arguments. 
A weak deformation of the globule, Figure~\ref{fig:globules}~a, into the ellipsoid, Figure~\ref{fig:globules}~b, 
produces a restoring force (\ref{eq:force_el}). The phase segregation inside the weakly deformed globule starts when this force becomes equal  or  comparable to the plateau force, Eq.~(\ref{f_c}) .
This gives the threshold extension
\begin{equation} \label{eq:TP1}
  D\simeq 2(\xi_t + R_{0}) \simeq 2[\xi_t + (Na^2\xi_t)^{1/3}] \simeq 2(Na^2\xi_t)^{1/3}
\end{equation}
which size is of the same order as that of the unperturbed globule.

\subsection{Tadpole-to-open chain transition}
The next step is to consider the transition from a microphase segregated tadpole, Figure~\ref{fig:globules}~c, to the open conformation, Figure~\ref{fig:globules}~d. In the transition point the free energies of the tadpole and open conformation should be equal: $F_{tadpole} = F_{chain}$. The free energy of the tadpole is given by Eq.~(\ref{eq:tadp:all}). The free energy of the extended (open) chain is calculated similarly to $F_{tail}$, Eq.~(\ref{eq:tadp:tail}). 
\begin{equation} \label{eq:chain:F}
  \frac{F_{chain}}{k_B T} \simeq \frac{D^2}{N a^2} \simeq \frac{(N-n)^2 a^2}{\xi_t^2 N} .
\end{equation}
where we have use the approximation that in the vicinity of the transition point the tail length is much larger than the globule size and, therefore, the tail size can be considered as being approximately equal to the overall chain extension
\begin{equation} 
  D \simeq \frac{N-n}{g_t}\, \xi_t \simeq \frac{(N-n) a^2}{\xi_t}
\end{equation}

Equating the free energies of the tadpole and the open chain, we obtain that the number of monomers in the head of the tadpole (globule) in the transition point scales as
\begin{equation} \label{eq:TP2:ngl}
  n \simeq N^{3/4} \left(\frac{\xi_t}{a}\right)^{1/2}
\end{equation}
or
\begin{equation}
 \frac{n}{g_t}\simeq \left(\frac{N}{g_t}\right)^{3/4}
\end{equation}
that is, the number of blobs in the ``minimal globule'' is determined by the number of blobs in the whole chain. Scaling dependence of $n$ vs. $N$ in the transition point with the chain length $N$ was obtained by Cooke and  Williams~\cite{Cooke:2003}, who considered the limiting case of a "dry" globule.

As a result, the size of the head at the transition point $R_{globule}\simeq (na^2\xi_t)^{1/3}\simeq N^{1/4} (\xi_t a)^{1/2}$ depends on $N$ and is much larger than the thermal blob size $\xi_t$. The jump in the reaction force is
\begin{equation} \label{eq:TP2:df}
     \frac{\Delta f}{k_B T} = \frac{f_{tadpole} - f_{chain}}{k_B T} \simeq 
     \frac{D}{(N-n) a^2}- \frac{D}{N a^2} \simeq
     \frac{1}{a}\cdot N^{-1/4} \left(\frac{\xi_t}{a}\right)^{-1/2}.
\end{equation}
We see that the jump in the force at the transition point should 
increase with an increase in $\chi$ and
decrease with an increase in $N$ thus vanishing in the thermodynamic limit $N\rightarrow \infty$ .

\subsection{Arbitrary $d$ case}
The above scaling analysis can be generalized in a way that takes into account an arbitrary dimensionality of the system $d$. 
In addition to the spherical globule case, which is the main subject of the present paper, corresponding to $d=3$, 
we consider a polymer ``bridging brush'' immersed into a poor solvent and pulled by an external force~\cite{Klushin:1998}, 
which is effectively a one-dimensional system, $d=1$. 
Similarly, as a two-dimensional case, $d=2$, we can propose a 
cylindrical brush (bottle-brush) whose arms are simultaneously and equally pulled out in the radial direction (out of main chain or the grafting line).

If the globular head contains $n$ monomers, the size of the globule scales as \\
$R_{globule} \sim \xi_t \left(na^2/\xi_t^2\right)^{1/d}$. Hence, the surface area per globule (per chain) is \\
$S_{globule}\sim \left(R_{globule}/\xi_t \right)^{d-1} \xi_t^2 \sim \left(n a^2/\xi_t^2 \right)^{(d-1)/d} \xi_t^2$. This will generalize the surface contribution to the free energy of the tadpole, the second term in Eq.~(\ref{eq:tadp:all}), the other contributions to $F_{tadpole}$ as well as to $F_{chain}$ remain unchanged. The relation between the tail length and $n$ is  $D-2R_{globule} \simeq (N-n)a^2/\xi_t$.

Taking this into account, the generalized expressions for the reaction force in the tadpole regime (\ref{eq:tadp:force}) reads
\begin{equation} \label{eq:tadp:force:arbd}
  \frac{f_{tadpole}}{k_B T} \simeq 
  \frac{1}{\xi_t}\left[1 - \frac{d-1}{d}\left(\frac{na^2}{\xi_t^2}\right)^{-1/d} \right] \simeq 
  \frac{1}{\xi_t}\left[1 - \frac{d-1}{d}\left(\frac{n}{g_t}\right)^{-1/d} \right]
\end{equation}
Since with an increase in $D$, $n$ obviously decreases. Eq.~(\ref{eq:tadp:force:arbd}) shows that $f_{tadpole}$ is a (weakly) decreasing function of the extension $D$ for $d = 2$ and 3, whereas for $d=1$ it is constant giving a true plateau on the force-extension curve.

For the location of the transition of the tadpole to the open chain the expression (\ref{eq:TP2:ngl}) for the number of monomers in the tadpole's head is modified as
\begin{equation} \label{eq:arbd:TP2:n}
  n \simeq N^{\frac{d}{d+1}} \left(\frac{\xi_t}{a}\right)^{\frac{2}{d+1}} \simeq 
  N^\frac{d}{d+1} g_t^\frac{1}{d+1}
\end{equation}
and for the force jump, Eq.~(\ref{eq:TP2:df}), we have
\begin{equation} \label{eq:arbd:TP2:df}
     \frac{\Delta f}{k_B T} \simeq \frac{1}{\xi_t}\cdot \left(\frac{\xi_t^2}{Na^2}\right)^{\frac{1}{d+1}} \simeq
     \frac{1}{\xi_t}\cdot \left(\frac{g_t}{N} \right)^\frac{1}{d+1}.
\end{equation}

Remark that in the case $d=1$ corresponding to the planar polymer brush we recover the scaling $n \sim N^{1/2}$ obtained earlier for the size of the microphase emerging in the brush capable to undergo a first order phase transition~\cite{Birshtein:2000, Klushin:2001, Birshtein:2003, Amoskov:2003}.

\subsection{Comparison of SCF results and scaling dependences}
The numerical results of the SCF modeling are in good qualitative agreement with the scaling laws obtained in the framework of the blob picture. In more detail, the result of the SCF theory have confirmed the existence of three regimes of extension of the globule as predicted by the blob model, namely (i) the deformation (extension) of the globule as a whole, (ii) the coexistence of a globular and unfolded (extended) phases, (iii) the disappearance of the globule and the further extension of the unfolded chain. The SCF results have further pointed to the possible departure from this scheme in the case of short chains and moderate values of the solvent quality $\chi$.

A qualitative agreement of results of the SCF calculations and the blob model is also found for the dependences of the shape of the force-extension curve on the solvent quality and chain length. 
The parameters of the blob model are: (i) the blob size which decreases with $\chi$ and (ii) the total number of blobs in the chain, the latter grows as a function of $N$ and $\chi$. According to the blob model, the formation of an extended tail starts when the extension slightly exceeds the unperturbed globule size  $D\sim (Na^2\xi_t)^{1/3}$. Correspondingly, the value of the ratio $D/(Na)$ at this transition point should decrease upon an increase in $\chi$ and $N$, exactly as is observed in the SCF modeling. According to the blob model, the height of the quasi-plateau on the force-extension curve should increase as a function of $\chi$ (due to the decrease in $\xi_t$) and $N$, that is, upon the increase in the total number of blobs. The latter determines the magnitude of the correction term in Eq.~(\ref{eq:tadp:force}). A decrease in the number of blobs in the globular phase, $n/g_t$, leads to a decreasing reaction force. Also, the jump-like decomposition of the globule at large $D/(Na)$, as predicted by the scaling analysis is in agreement with the SCF analysis.

At this stage it is necessary to realize that the blob model is not strictly applicable for the values of 
$\chi$ and $N$ used in our SF-SCF simulations. The notion of a Gaussian thermal blob itself only has a meaning when the number of the monomer units 
in the blob, $g_t$, is large enough. 
This is only the case at relatively small deviation from the $\Theta$-point, $\tau=(\Theta-T)/T\ll 1$. 
The range of $\chi$-values considered in the present work obviously does not satisfy this demand. 
Furthermore, the total number of thermal blobs in the chain is $\sim N\tau^2$ and this number should be large to use scaling arguments. 
Hence, the scaling parameter of the blob model is $N^{1/2}\tau$, that requires at small values of $\tau$ much larger chain lengths $N$ 
as compared to those used in the numerical analysis. In the SCF modeling the value of $N$ used in the calculations was restricted from above by computational reasons. 

Admittedly, the blob model includes some simplifying assumptions as well. In particular, it is assumed that in the tadpole conformation, the globular head has a spherical shape. However, as one can see from the density profiles, Figure~\ref{fig:profiles_200_chi14}, the head has a prolate (ellipsoidal) shape rather than a spherical one. As a consequence, in the prolate globule, there appears a reaction force tending to restore its unperturbed spherical shape. In the tadpole this force is balanced by the elastic force from the tail. The analytical model that takes into account the non-spherical shape of the tadpole head and properly accounts for the force equilibrium will be considered in a follow up publication. We note, however, that the ``spherical head assumption'' does not affect the scaling dependences obtained in this section.

\section{Conclusions}

We have performed detailed SCF calculations of the equilibrium unfolding of a globule formed by a flexible homopolymer chain collapsed in a poor solvent 
and subjected to an extensional deformation.  
More specifically, we consider the conformational characteristics adopted by a chain with imposed end-to-end distances 
in a wide range of (poor) solvent qualities (expressed in terms of Flory-Huggins solubility parameter $\chi$) and polymerization degrees $N$. 
The fluctuating restoring force (i.e., the elastic force) is calculated as a function of the end-to-end distance. 
These results are collected in force-deformation curves.

In accordance to predictions of the scaling theory by Halperin and Zhulina, 
we have observed a sequence of intra-molecular conformational transitions that occur upon an increase of the deformation.  
We have found that there is a linear response regime (found at small deformations). 
This is followed by an intra-molecular  micro-phase segregation regime (found at intermediate extension): 
here a uniformly stretched segment of the chain (a "tail") co-exists with a collapsed globular domain (a ``core''). 
A progressive increase in the end-to-end distance of the chain is accompanied by a systematic depletion of the globular 
core and a re-partitioning of the monomer units into the stretched tail. 
We have found that the unfolding of the globular core occurs at a weakly decreasing (fairly constant) reaction force,  
whereas the entropic elasticity is predicted to be recovered at strong extensions.

The general shape of the force-extension curves obtained using the SF-SCF approach appears to be unconventional, showing a more or less extended region with an anomalous dependence with $df/dD < 0$, i.e. with a negative extensional modulus. In fact, $f(D)$ curves exhibit the van der Waals loop, which is usually an indicator of the instability of a system. However, if the role of the governing parameter is played by the end-to-end distance, the system cannot avoid these states and passes through them step by step thus undergoing a sequence of intra-molecular conformational transitions.

We have found that the intra-molecular co-existence occurs only 
for sufficiently long polymers and at large values of the $\chi$ parameter. 
In other cases, that is, for relatively short chains (number of segments of the order of $10^2$) and mild 
solvent conditions $\chi \sim 1$, a uniformly stretched conformation is retained and no microphase segregation is predicted 
to occur in the whole range of extensional deformations. 
Nevertheless the force-extension curve exhibits a region with $df/dD < 0$. Hence, we predict that the phase diagram of the system in $N,D$ or $\chi,D$ coordinates contains an one-phase and two-phase regions and thus exhibits a critical point.

Furthermore, we have analyzed two conformational transitions: (i) the first one is from the 
weakly elongated (ellipsoidal) globule to the tadpole, which consists of a ellipsoidal globular core coexisting with a stretched tail and (ii) the second one is from the tadpole conformation to that of a uniformly stretched chain --  the unraveling transition discovered by Cooke and Williams~\cite{Cooke:2003}. The first transition, that is the formation of a stretched intra-molecular micro-phase occurs continuously, whereas the second transition occurs as the first order phase transition and the reaction force drops abruptly down at certain elongation threshold. 

We anticipate that the patterns predicted by our theory should manifest in force-deformation spectra that can be obtained, e.g., by means of single-molecule AFM spectroscopy on end-grafted polymers
collapsed upon  a decrease in the solvent strength. The latter can be achieved by variation in temperature or in pH for thermo- or pH sensitive polymers, respectively. The most straightforward way for a synthetic implementation of this system is to graft polymer chains to a solid surface, for example by means of radical polymerization initiated at the surface. Typical degrees of polymerization obtained in controlled radical polymerization approach values as high as $10^2 - 10^3$. 

It is interesting to point to other polymeric systems that behave similarly to the stretched globule in a sense that these relieve the stress caused by an external field by ``throwing out'' a part of the chain as a new phase, thus forming a microphase segregated state. This is for example the case for a Gaussian chain compressed between two pistons. This chain undergoes an abrupt transition from a confined coil state to an inhomogeneous flower-like conformation partially escaped from the gap~\cite{Klushin:2004, Skvortsov:2007}. In the transition (microphase coexistence) region this system exhibits a negative compressibility, i.e. the reaction force decreases with an increase in deformation (corresponding to a decrease in a distance between the pistons, or chain squeezing) whereas at small and large deformation it grows. 


In the presented paper all results concern the deformation of a globule in the constant extension ensemble. 
In the conjugate constant force ensemble, one should expect that the tadpole conformation will be unstable and the globule will unfold jumpwise, as a transition going from the ellipsoid globule directly to the open chain. This is confirmed by the results of molecular dynamics~\cite{Frisch:2002} and Monte Carlo~\cite{Grassberger:2002} simulations. However, in the case of lattice models or when the formation of helical conformation is possible a multistep transition can be observed~\cite{Marenduzzo:2003, Marenduzzo:2004}.

\section*{Acknowlegdement}
\label{Acknowlegdement}
This work has been performed as a part of
the collaborative research project SONS-AMPHI within the European
Science Foundation EUROCORES Program, and has been
partially supported by funds from the EC Sixth Framework Program
through the Marie Curie Research and Training Network POLYAMPHI.
Support by the Dutch National Science Foundation
(NOW) and the Russian Foundation for Basic Research (RFBR)
through Joint Project 047.017.026/06.04.89402 and Project 08-03-00336a
is gratefully acknowledged.

\appendix

\section{Conformational free energy of (strongly) stretched chains on a cylindrical lattice}\label{app:Fconf}
The aim of this Appendix is to derive the conformational free energy of a chain in a cylindrical lattice as a function of the end-to-end distance $D$. 
It can be easily calculated if we consider an ideal chain extended by an external force (external field), i.e. we temporarily ``switch'' 
from the fixed stretching ensemble studied in this paper to the fixed force ensemble, 
that is the common Ansatz in statistical physics, and then switch back to the fixed deformation ensemble studied in this work. 

Suppose that the chain walking on the cylindrical lattice is subjected to a force $f $ directed along the $z$-axis. For the sake of comparison with numerical results, we set the transition probabilities for the random walk according to SF-SCF model. Namely, the probability to make a step either in $r$ or in $z$ direction is given by $\lambda_1$, the probability of a step in ``$rz$''-direction, i.e. simultaneously changing both $r$ and $z$ coordinates by $\pm 1$ is $\lambda_2$ whereas the rest, $\lambda_0=1-4\lambda_1-4\lambda_2$ is the probability to change the angular coordinate $\varphi$. Then the statistical weight of a monomer (of a link) is given by
\begin{equation}  \label{eq:wchain_f}
  \begin{split}
    w  & = (\lambda_0 + 2\lambda_1) e^0 + (\lambda_1 + 2\lambda_2) e^{f a/k_BT} + (\lambda_1 + 2\lambda_2) 
    e^{-f a/k_BT} \\ 
    & = 1-2\lambda_1 - 4\lambda_2 + 2(\lambda_1 + 2\lambda_2)\cosh\left(\frac{f a}{k_BT}\right)
  \end{split}
\end{equation}
Then the partition function of the chain is
\begin{equation}  \label{eq:Zchain_f}
  Z  = \left[1-2\lambda_1 - 4\lambda_2 + 2(\lambda_1 + 2\lambda_2)\cosh\left(\frac{f a}{k_BT}\right)\right]^N
\end{equation}
The logarithm of the partition function gives us \emph{the Gibbs free energy}, 
\begin{equation}  \label{eq:Fchain_f}
  G_f = - k_B T \log Z  = -Nk_B T\cdot \log \left[1-2\lambda_1 - 4\lambda_2 + 2(\lambda_1 + 2\lambda_2)\cosh\left(\frac{f a}{k_BT}\right)\right]
\end{equation}
Once the partition function (the Gibbs free energy) is known, the extension $D$ corresponding to the force $f $ can be found straightforwardly:
\begin{equation}  \label{eq:Dchain_f}
  D = - k_B T \cdot \frac{\partial G_f}{\partial f }  = 
  Na\cdot \frac{2(\lambda_1 + 2\lambda_2)\sinh\left(\frac{f a}{k_BT}\right)}{1+ 
  2(\lambda_1 + 2\lambda_2) \left[\cosh\left(\frac{f a}{k_BT}\right) - 1\right]}
\end{equation}
Eq.~(\ref{eq:Dchain_f}) gives the dependence of the \emph{reduced} extension (or the degree of stretching) $D/(Na)$ on the applied force. This equation 

and we see that the force is a universal function of the \emph{reduced} extension $D/(Na)$. 

In order to return to the fixed extension ensemble, one can use the standard relation between the free energies in $f$- and $D$- ensembles (Gibbs and \emph{Helmholz} free energies, respectively)
\begin{equation} 
  F_D = G_f + D\cdot f
\end{equation}
this gives the conformational free energy of the chain having its ends fixed at the distance $D$:
\begin{equation}  \label{eq:Fchain_D}
  \begin{split}
    F_{chain}=F_D = & -Nk_B T\cdot \log \left[1-2\lambda_1 - 4\lambda_2 + 2(\lambda_1 + 2\lambda_2)\cosh\left(\frac{f a}{k_BT}\right)\right] + \\
    & Naf\cdot \frac{2(\lambda_1 + 2\lambda_2)\sinh\left(\frac{f a}{k_BT}\right)}{1+ 
    2(\lambda_1 + 2\lambda_2) \left[\cosh\left(\frac{f a}{k_BT}\right) - 1\right]}
  \end{split}
\end{equation}
In Eq.~(\ref{eq:Fchain_D}) $F$ is expressed as a function of $f$ but together with Eq.~(\ref{eq:Dchain_f}) it parametrically defines $F$ as a function of $D$. 

In the weak deformation limit we obtain
\begin{equation}  \label{eq:Fchain_appr}
  \frac{F_{chain}}{k_B T} =\frac{1}{4(\lambda_1 + 2\lambda_2)} \cdot \frac{D^2}{Na^2}.
\end{equation}
So, we see that the elastic free energy has the Gaussian form
\begin{equation}
	\frac{F_{chain}}{k_B T} = k \cdot \frac{D^2}{Na^2}
\end{equation}
with  $k =1/4(\lambda_1 + 2\lambda_2)$. The corresponding reaction force is
\begin{equation}  \label{eq:force_chain_appr}
  \frac{f_{chain}}{k_B T} = 2 k \cdot \frac{D}{Na^2}.
\end{equation}

In the strong deformation limit $\sinh(fa/k_BT)\simeq \cosh(fa/k_BT)\simeq \frac{1}{2}\exp(fa/k_BT)$ and  Eq.~(\ref{eq:Dchain_f}) gives
\begin{equation}  \label{eq:force_chain_strong}
  \frac{f_{chain}}{k_B T} = \log\left[\frac{2(2k-1)\cdot D/(Na)}{1- D/(Na)} \right].
\end{equation}
The reaction force asymptotically tends to infinity as $D$ approaches $Na$.

Comparing analytical force-extension curve (\ref{eq:Dchain_f}) with SF-SCF numerical results (Figures~\ref{fig:force_200} and \ref{fig:force_all}; analytical curves are shown by dotted lines), we see a perfect correspondence in the uniformly stretched chain regime.


\end{document}